\documentclass[reprint,aps,prb,amsmath,amssymb,longbibliography, notitlepage]{revtex4-1}
\usepackage{amsmath,amssymb,bm,graphicx}
\usepackage{lipsum}
\usepackage{graphics}
\usepackage{algorithmic}
\usepackage[stable]{footmisc}
\usepackage{soul}
\usepackage{physics}
\usepackage{pifont}
\usepackage{mathrsfs}
\usepackage{mathalpha} 
\usepackage{amstext}
\usepackage[version=4]{mhchem}
\usepackage{cancel}
\usepackage{makecell}
\usepackage[normalem]{ulem}
\RequirePackage{silence}
\usepackage[colorlinks, linkcolor= blue, citecolor = blue, urlcolor=blue]{hyperref}
\newcommand{\cmark}{\ding{51}}%
\newcommand{\xmark}{\ding{55}}%
\def\be{\begin{equation}}
\def\ee{\end{equation}}
\def \bea{\begin{eqnarray}}
\def \eea{\end{eqnarray}}
\def \nn{\nonumber}
\def \cal{\mathcal}
\def \rm{\mathrm{}}

\begin{document}

\title{Intrinsic Gyrotropic Magnetic Current of Orbital Origin}

\author{Koushik Ghorai}
\email{koushikgh20@iitk.ac.in}
\author{Sankar Sarkar}
\email{sankars24@iitk.ac.in}
\author{Amit Agarwal}
\email{amitag@iitk.ac.in}
\affiliation{Department of Physics, Indian Institute of Technology Kanpur, Kanpur-208016, India}

\begin{abstract}
In gyrotropic crystals, an oscillating magnetic field induces a charge response known as the gyrotropic magnetic current. While its conventional origin is attributed to magnetic field modified band energy and shift in the Fermi-surface, a recent study identified an additional spin-driven magnetic displacement contribution. Here, we complete the picture by identifying the orbital counterpart of the magnetic displacement current. Using a density-matrix formulation that incorporates both minimal coupling and spin–Zeeman interactions, we derive the electronic equations of motion in the presence of an oscillating magnetic field and uncover a previously unexplored orbital contribution to the wavepacket velocity. Physically, this contribution arises from the time variation of the magnetic field induced charge polarization. In the low frequency transport regime, this mechanism becomes purely intrinsic. We illustrate this intrinsic gyrotropic current of orbital origin in the ${\cal P}{\cal T}$-symmetric antiferromagnet $\ce{CuMnAs}$. We show that the intrinsic gyrotropic magnetic current reverses sign upon N\'eel vector reversal, establishing it as a direct probe of antiferromagnetic order in $\ce{CuMnAs}$ and other $\mathcal{PT}$-symmetric antiferromagnets.

\end{abstract}

\maketitle

\section{Introduction}

The gyrotropic effect typically refers to the generation of an axial (polar) response for a polar (axial) driving field. It includes a variety of phenomena such as optical activity~\cite{Landau,Burkov_1975,Moore_15_prl}, circular dichroism~\cite{Hosur_15_prb}, Faraday rotation, circular photogalvanic effect~\cite{Rogachev_1979_ssc}, magnetoelectric effect~\cite{KTLaw_20_prr,Sayantika_24_prr,Sayan_25_njp}, and gyrotropic Hall effect~\cite{Pesin_19_prb} amongst others. A particularly intriguing effect is the gyrotropic magnetic current (GMC), where a slowly varying magnetic field drives a dissipationless electrical current. GMC in metals was first identified by Zhong \emph{et al.} as a low-frequency limit of optical gyrotropy~\cite{Moore_2016_prl}. It has been explored in Weyl semimetals~\cite{Sumanta_15_prb} and chiral magnets~\cite{Paul_25_prl}. Since GMC arises from the magnetic moments of quasiparticles near the Fermi surface, it has become a sensitive probe of materials' symmetry, Berry curvature, degeneracy of chiral fermions, and band geometric quantities~\cite{Nandy_20_scirep,Amit_24_review, Grushin_18_prb, Atasi_22_NatPhys,Atasi_22_2dMat}.

Microscopically, GMC originates from intrinsic and disorder-induced corrections to the electronic magnetic moment~\cite{Pesin_17_prb}. Recent works have pointed out that beyond the Zeeman-coupling induced modification of band energies, magnetic driving induces a positional shift of the electronic wavepacket, leading to a finite charge polarization~\cite{Niu_prl_14,Gao_19_prl}. For an oscillating magnetic field, the temporal modulation of this polarization generates a magnetic analogue of the electric field induced displacement current~\cite{Jian_Wang_2024_PRB,Tobias_2024_NatComm}. Recently, Wang \emph{et al.}~\cite{Wang_2025_prl} identified the spin contribution to this magnetic displacement current and demonstrated that it constitutes an intrinsic gyrotropic magnetic current (IGMC) at finite frequency. However, the corresponding orbital contribution was not addressed. This omission is particularly significant in view of growing evidence that orbital mechanisms dominate a wide range of responses in weak spin–orbit coupled materials~\cite{HyunLee_23_nature,Oppeneer_23_prl,Pietro_23_prl,DongGo_25,2DPHE_koushik,Yamakage_25_arxiv,ShenYang_2025_prb,Rahul_25_prb}. Orbital effects have also been shown to generate additional contributions to electronic charge transport~\cite{Kamal_21_prb,Amit_24_review,Sunit_25_prb,WenHe_25_prr,Carmine_25_arxiv}.

%%%%%%%%%%%%%%%%%%%%%%%%%%%%%%%%%%%%%%%%%%%%%%%%%%%%%%%%%%%
\begin{figure}[t!]
    \centering
    \includegraphics[width=.9\linewidth]{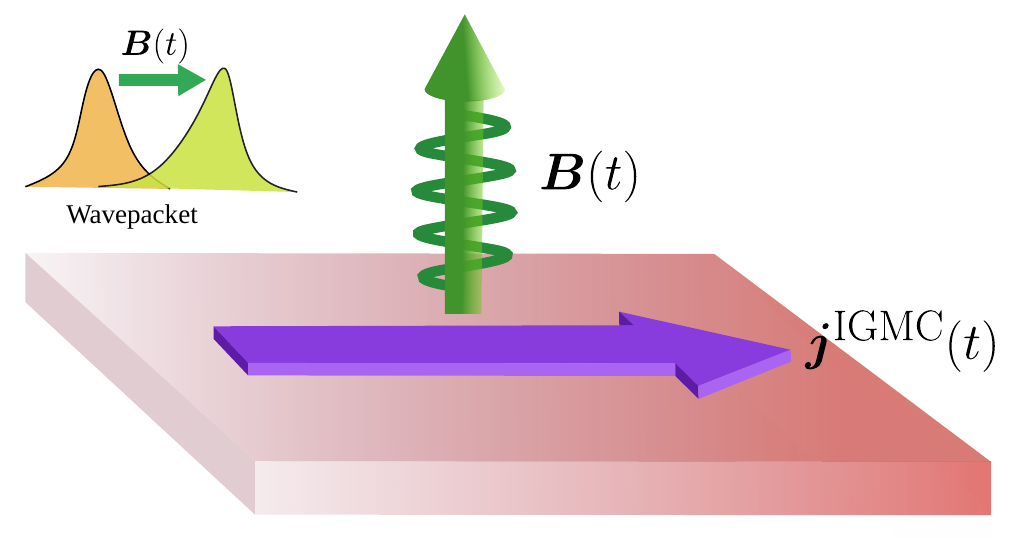}
    \caption{Schematic of the intrinsic gyrotropic magnetic current (IGMC) of orbital origin. A time-varying magnetic field $\bm{B}(t)$ couples to the electronic orbital magnetic moment, inducing an orbital contribution to the oscillatory positional shift of the wave packet and dynamic charge polarization. The time variation of this polarization gives rise to the IGMC response.}
    \label{Fig_Schematic}
\end{figure}
%%%%%%%%%%%%%%%%%%%%%%%%%%%%%%%%%%%%%%%%%%%%%%%%%%%%%%%%%%%

In this work, we develop a unified theory of IGMC that treats spin and orbital coupling on an equal footing. Starting from a density-matrix framework, we derive the semiclassical, finite-frequency equations of motion (EOM) of electron wavepackets in the presence of an oscillating magnetic field.  This enables us to identify the previously overlooked orbital contribution to IGMC arising from the orbital-induced positional shift of the Bloch electrons. We quantify this positional shift, or the Berry connection correction, by introducing a band geometric quantity: \emph{orbital}-magnetic Berry connection polarizability (MBCP). Its spin counterpart is denoted as \emph{spin}-MBCP. Importantly, this density matrix formulation of Berry connection correction contains an extra multiband term in MBCP that was missing in earlier semiclassical treatments~\cite{Niu_prl_14,ShenYang_2025_prb}.

In contrast to the Fermi surface nature of conventional GMC, IGMC is a Fermi sea response, and it is finite in both metals and in insulators. Another interesting distinction between them is that while IGMC is symmetry allowed and becomes the leading gyrotropic response in systems with combined inversion ($\mathcal P$) and time-reversal ($\mathcal T$) symmetry (bipartite antiferromagnets), conventional GMC is restricted in such systems by symmetry. We further analyze the crystallographic point group symmetry of the IGMC response tensors to facilitate the right material selection. Owing to this, we demonstrate the orbital-driven IGMC in ${\cal P}{\cal T}$-symmetric antiferromagnetic phase of tetragonal $\ce{CuMnAs}$. Notably, the orbital mechanism of IGMC dominates the response and is almost an order of magnitude larger than its spin counterpart in CuMnAs. Being a time-reversal-odd response, IGMC shows a distinct signature in $\ce{CuMnAs}$: it reverses sign as the magnetization order, or the N\'eel vector, is reversed. This establishes IGMC as an interesting mechanism to probe the antiferromagnetic ordering.

The rest of the paper is organized as follows: Section \ref{Sec: Density Matrix} develops the density matrix formulation of the finite frequency EOM. Section \ref{Sec: GMC} presents different contributions to IGMC, including the novel contribution due to the orbital part of the magnetic displacement current. We discuss the fundamental and crystallographic point group symmetry restrictions of the IGMC responses in Section \ref{Sec: Symmetry}. We demonstrate the IMGC response in antiferromagnetic $\ce{CuMnAs}$ in Section \ref{Sec: CuMnAs}. In Section \ref{Sec: Conclusion}, we summarize our main findings.

%%%%%%%%%%%%%%%%%%%%%%%%%%%%%%%%%%%%%%%%%%%%%%%%%%%%%%%%%%%%%%%%%

\section{Density matrix formulation of equation of motion} \label{Sec: Density Matrix}

In this section, we derive the electronic EOM in the presence of a harmonically oscillating magnetic field of amplitude $\bm{B}_0$ and frequency $\omega$,
\be
\bm{B}(t) = \tfrac{1}{2}\bm{B}_0 e^{i\omega t} + \text{c.c.}
\ee
We derive the semiclassical dynamics of Bloch electrons in crystalline materials using the Euler-Lagrange formalism. Semiclassically, the Lagrangian is obtained by taking the expectation value of the Lagrangian operator $\hat{\cal{L}} = i\hbar \partial_t - \hat{\mathcal{H}}$ over the electronic wavepacket $\ket{W}$,
\bea
\mathcal{L} = \bra{W}\, i\hbar \frac{\partial}{\partial t} - \hat{\cal{H}} \,\ket{W}~.
\eea
Here, $\hat{\cal H} = \hat{\cal{H}}_0 + \hat{\cal{H}}^{B}$ is the total Hamiltonian with an unperturbed $\hat{\cal H}_0$ and the magnetic interaction term $\hat{\cal H}^B$. In this work, we consider the magnetic coupling with both orbital and spin magnetic moments, giving rise to the total magnetic perturbation
\bea
\hat{\mathcal{H}}^{B} = \hat{\mathcal{H}}^{B}_{L} + \hat{\mathcal{H}}^{B}_{S}
= -\,\bm{B}\!\cdot\!\hat{\bm{\mathcal{M}}}^{L} - \bm{B}\!\cdot\!\hat{\bm{\mathcal{M}}}^{S}~.
\eea
Here, $\hat{\bm{\mathcal{M}}}^{L} = -(e/4)[\hat{\bm{r}} \times \hat{\bm{v}} - \hat{\bm{v}} \times \hat{\bm{r}}]$ is the orbital magnetic-moment operator, and $\hat{\bm{\mathcal{M}}}^{S} = -(\mathsf{g}_s\mu_B/2)\hat{\bm{\sigma}}$ is its spin counterpart. 
Here, $\hat{\bm{r}}$ and $\hat{\bm{v}}$ refers to the electronic position and velocity operators, respectively, $\mathsf{g}_s \simeq 2$ is the spin $g$-factor, $\mu_B$ denotes the Bohr magneton, and $\hat{\bm{\sigma}}$ corresponds to the Pauli matrices.

Instead of following the conventional wavepacket formalism~\cite{Sundaram_1999_prb, Gao_19_prl}, here we employ a density-matrix approach~\cite{Boyd_NonlinearOptics} to construct the Lagrangian. The advantage of this formulation is twofold: first, it unifies these two frameworks under dynamic magnetic perturbation; second, the finite frequency density matrix solution can be employed in other magnetotransport studies. In the density matrix formulation, the system Lagrangian is defined as the trace of the Lagrangian operator $\hat{\mathcal{L}}$ weighted by the density matrix $\hat{\rho}$, 
\bea \label{def_Lag}
\mathcal{L} = \mathrm{Tr}\{\hat{\mathcal{L}}\hat{\rho}\}~.
\eea

In the presence of a magnetic field, the evolution of the density matrix is described by the quantum Liouville equation~\cite{Dimi_PRB_17}. We have,  
\begin{equation} 
\frac{\partial \hat{\rho}}{\partial t} + \frac{i}{\hbar}\comm{\hat{\mathcal{H}}_0}{\hat {\rho}} + \frac{1}{\tau}\hat{\rho} = {\mathcal D}_B \hat{\rho} - \frac{i\mathsf{g}_s \mu_B}{2\hbar} \bm{B} \cdot [\hat{\bm{\sigma}}, \hat{\rho}]~,\label{QKT_B}
\end{equation}
where $\tau$ denotes the scattering time that captures the relaxation of the electronic states of the system. The first term on the right-hand side describes the coupling between the magnetic field and the orbital motion of Bloch electrons, while the second represents the Zeeman coupling to spin. The explicit form of the orbital contribution is given by
\bea
[\mathcal{D}_B\hat{\rho}]_{nm} = \frac{e}{2\hbar^2} B_0^c \epsilon_{cab} \left\{ {\mathcal D}_b\hat{\mathcal{H}}_0, {\mathcal D}_{a} \hat{\rho}\right\}_{nm}~,
\eea
where $\{a,b,c\}$ denote Cartesian components, and summation over repeated indices is implied. The action of the covariant derivative~\cite{Sipe_95_prb} in momentum space on an operator $\hat{\cal{X}}$ is given by
\bea
[{\mathcal D}_{\bm k} \hat{\mathcal{X}}]_{nm} \equiv \left(\frac{{\mathcal D} \hat{\mathcal{X}}}{{\mathcal D}\bm{k}}\right)_{nm} = \partial_{\bm k}\mathcal{X}_{nm} - i[\bm{\hat{\mathcal{R}}},\hat{\mathcal{X}}]_{nm}~.
\eea
Here, $[\bm{\hat{\mathcal{R}}}]_{nm} \equiv \bm{\mathcal{R}}_{nm} = \bra{u_{n\bm k}}i\partial_{\bm k}\ket{u_{m\bm{k}}}$ is the band-resolved Berry connection, and $\ket{u_{n\bm k}}$ is the cell-periodic part of the Bloch eigenstates $\ket{\psi_{n\bm k}} = e^{i\bm{k}\cdot \bm{r}}\ket{u_{n\bm k}}$ for an $n^{\mathrm{th}}$ band electron with momentum $\bm{k}$ and eigenvalue $\varepsilon_{n\bm k}$. The spin part of the magnetic interaction is captured by the second term on the right-hand side in Eq.~(\ref{QKT_B}).

In the weak field limit, we solve Eq.~(\ref{QKT_B}) perturbatively to obtain the density matrix elements at different orders in magnetic field strength. The first-order solution of the density matrix is expressed as, 
\be \label{rho_t}
\rho_{nm}^{B}(t) =  {\rho}_{nm}^{B}(\omega)e^{i\omega t} +  {\rho}_{nm}^{B}(-\omega)e^{-i\omega t}~.
\ee
The harmonic components of the density matrix can be segregated into an orbital ($\rho_{nm;L}^{B}$) and a spin part ($\rho_{nm;S}^{B}$),
\bea
\rho_{nm}^{B} (\omega) &=& \rho_{nm;L}^{B} (\omega) + \rho_{nm;S}^{B} (\omega)~.
\eea
We calculate these terms explicitly in Appendix. \ref{First_order_DM} to obtain, 
\bea \label{rho_nm_B_v2}
\rho_{nm;S}^B (\omega) = \frac{i\mathsf{g}_s \mu_B}{4\hbar} g^{\omega}_{nm} [\bm{B}_0\cdot \bm{\sigma}_{nm}]f_{nm}~,
\eea 
and 
\begin{widetext}
\bea 
\rho_{nm;L}^B (\omega) &=& \frac{e}{4\hbar}g^{\omega}_{nm} \bm{B}_0 \cdot \left[ \{\partial_{\bm k}(f_n+f_m)\} \times \bm{v}_{nm} \right] +\frac{i}{2\hbar}g^{\omega}_{nm} \bm{B}_0 \cdot \sum_{p \neq m}\left[\bm{\cal M}_{npm}^Lf_{pm} - \left( \bm{\mathcal{M}}_{mpn}^{L} \right)^{*}f_{pn} \right]~.
\eea
\end{widetext}
Here  $g^{\omega}_{nm} = [1/\tau+i(\omega_{nm} + \omega)]^{-1}$ with $\hbar \omega_{nm} = \varepsilon_{n{\bm k}} - \varepsilon_{m{\bm k}}$ being interband transition frequency. 
Additionally, $f_{pm} = (f_p - f_m)$ is the equilibrium population difference, $\bm{\cal{M}}_{mpn}^L = -(e/2)(\bm{v}_{mp} + \delta_{mp}\bm{v}_n)\times \bm{\cal{R}}_{pn}$, and $\bm{v}_{nm} = (1/\hbar)\bra{u_{n\bm k}}\partial_{\bm k}{\cal H}_0 \ket{u_{m \bm k}}$ corresponds to the unperturbed velocity matrix element. We substitute this density matrix in Eq.~(\ref{def_Lag}) to obtain the $n^{\mathrm{th}}$ band Lagrangian, 
\be
\mathcal{L}_{n\bm k} = 
- \hbar\dot{\bm{k}} \cdot 
\left(\bm{\mathfrak{r}}_n - \tilde{\bm{\mathcal R}}_{n\bm{k}}\right)
+ \frac{e}{2}(\bm{B}\times \dot{\bm{\mathfrak r}}_n) \cdot \bm{\mathfrak{r}}_n 
- \tilde{\varepsilon}_{n\bm k}~, \label{lagrangian_n}
\ee
where ${\bm{\mathfrak r}}_n$ is the average electronic position in a unit cell, or, equivalently, the center of a semiclassical wavepacket. $\bm{~\tilde{\mathcal R}}_{n\bm{k}} = \bm{\cal R}_{n\bm{k}} + \bm{\mathcal{R}}_{n\bm{k}}^{B}(t)$ is the magnetic field induced total Berry connection, and $\tilde{\varepsilon}_{n\bm{k}} = {\varepsilon}_{n\bm{k}} - (\bm{\cal M}^L_{n\bm{k}} + \bm{\cal M}^S_{n\bm{k}})\cdot \bm{B}$ corresponds to the field-corrected band energy. The details of the calculation are presented in Appendix~\ref{Appendix B: Lagrangian}. 

The Berry connection correction ${\bm{\mathcal R}}^{B}_{n\bm{k}}(t)$ emerges due to the magnetic field driven shift in charge distribution and can be expressed in the following form
\bea
{\cal R}_{n\bm{k}}^{B;a}(t) &=& \tilde{\cal{G}}_{n\bm{k}}^{B;ad}(\omega)B_0^d e^{i\omega t} + \text{c.c.}
\eea
Here, we have introduced a band-geometric quantity $\tilde{\cal{G}}_{n\bm{k}}^{B;ad}(\omega)$ and we term it as magnetic Berry connection polarizability (MBCP). Similar to the electric field induced Berry connection polarizability~\cite{Niu_prl_14,Kamal_23_prb,Koushik_24_prl}, the magnetic field driven correction is also a result of field induced interband transitions or mixing of Bloch states and would vanish if the magnetic driving is switched off ($\bm{B}\rightarrow 0$). Since the interband hybridization can be generated by both minimal coupling perturbation and Zeeman interaction, both orbital and spin magnetic moments would contribute to the Berry connection correction. Accordingly, we split finite-frequency MBCP into a spin and an orbital part, 
\bea
\tilde{\cal{G}}_{n\bm{k}}^{B;ad}(\omega) = \tilde{\cal{G}}_{n\bm k;S}^{B;ad}(\omega) + \tilde{\cal{G}}_{n\bm k;L}^{B;ad}(\omega)~.
\eea
The \emph{spin}-MBCP $(\tilde{\cal{G}}_{n;S}^{B;ad})$ and \emph{orbital}-MBCP $(\tilde{\cal{G}}_{n;L}^{B;ad})$ are defined as follows: 
\bea
\tilde{\cal{G}}_{n\bm k;S/L}^{B;ad}(\omega) &=& {\cal{G}}_{n\bm k;S/L}^{B;ad}(\omega) + \left[{\cal{G}}_{n\bm k;S/L}^{B;ad}(-\omega)\right]^*~,\label{MBCP}
\eea
where
\bea \label{Mag_Berry_Connection_OMM_t}
{\cal{G}}_{n\bm k;S}^{B;ad}(\omega) &=& \frac{i}{2\hbar} \sum_{m\neq n} g^{\omega}_{mn} {\cal{M}}_{mn}^{S;d} {\mathcal R}_{nm}^a~, \label{spin_BCP}
\eea 
and 
\begin{widetext}
\bea 
{\cal{G}}_{n\bm k;L}^{B;ad}(\omega) &=& -\frac{e}{4\hbar}  \sum_{m\neq n}  \epsilon_{dbc}\partial_b  \biggl[g^{\omega}_{nm} v_{nm}^c{\mathcal R}_{mn}^a\biggr]  - \frac{i}{2\hbar} \sum_{m\neq n} g^{\omega}_{mn} {\cal{M}}_{mn}^{L;d}{\mathcal R}_{nm}^a  + \frac{i}{2\hbar} \sum_{m\neq p}\sum_{p \neq n} g^{\omega}_{mp}\cal{M}_{mnp}^{L;d} {\mathcal R}_{pm}^a~. \label{Orbital_BCP}
\eea
\end{widetext}%
Here, $\bm{\cal{M}}_{mn}^L = \sum_{p\neq n}\bm{\cal{M}}_{mpn}^L$ is the matrix element of the orbital magnetic moment operator. The \emph{spin}-MBCP has been explored in some recent articles~\cite{Shengyuan_2023_prl,Wang_2025_prl,Neelanjan_25} and it is also referred to as anomalous spin polarizability~\cite{Shengyuan_2024_prl}. The \emph{orbital}-MBCP emerges naturally from the present formulation. 
The first two terms in Eq.~(\ref{Orbital_BCP}) generalize the earlier result of the orbital contribution to the magnetic field induced  Berry connection~\cite{Gao_frontiers_19,Shengyuan_2024_prl} to include finite frequency and disorder effects (See Appendix~\ref{Appendix B: Lagrangian} for details). Additionally, the last term in Eq.~(\ref{Orbital_BCP}) captures a unique multiband contribution to the MBCP that has not been explored earlier.

In two-dimensional materials, the out-of-plane orbital magnetic moment can couple to only an out-of-plane magnetic field. Thereby, \emph{orbital}-MBCP would vanish for an in-plane magnetic field. However, the \emph{spin}-MBCP may survive in such a scenario. More importantly, the time-dependent Berry connection remains gauge-invariant, similar to its dc counterpart.

\subsection{Equation of Motion}

The dynamics of Bloch electrons can be obtained from the derived Lagrangian in Eq.~(\ref{lagrangian_n}) using Euler-Lagrange equations
\bea 
\frac{d}{dt} \left(\frac{\partial \mathcal{L}_{n \bm k}}{\partial \dot{\bm{k}}} \right) &=& \frac{\partial \mathcal{L}_{n \bm k}}{\partial \bm{k}}~,\\
\quad \frac{d}{dt} \left( \frac{\partial \mathcal{L}_{n \bm k}}{\partial \dot{\bm{\mathfrak{r}}}_{n \bm k}} \right) &=& \frac{\partial \mathcal{L}_{n \bm k}}{\partial \bm{\mathfrak{r}}_{n\bm k}}~.
\eea
The first Euler-Lagrange equation yields the quasiparticle velocity
\bea \label{Vel_coupled}
\dot{\bm{\mathfrak{r}}}_{n \bm k} &=& \tilde{\bm{v}}_{n\bm{k}} - \dot{\bm{k}} \times ( \bm{\Omega}_{n\bm{k}} + \bm{\Omega}_{n\bm{k}}^B ) + \frac{\partial \bm{\cal R}_{n\bm{k}}^{B}}{\partial t}~.
\eea
Here, $\tilde{\bm v}_{n\bm{k}} = (1/\hbar)\partial_{\bm k} \tilde{\varepsilon}_{n\bm k}$ is the magnetic field corrected band velocity and $ \bm{\Omega}_{n\bm k} ( \bm{\Omega}_{n\bm k}^B) = \partial_{\bm k}\times \bm{\cal R}_{n\bm{k}} (\bm{\cal R}_{n\bm{k}}^B)$ is the unperturbed (magnetic field induced) Berry curvature of the system.
The second Euler-Lagrange equation produces the familiar force equation,
\bea \label{Force_coupled}
\hbar \dot{\bm{k}} &=& -e[\dot{\bm{\mathfrak{r}}}_{n \bm k} \times \bm{B}(t)]~.
\eea
These equations of motion can be decoupled to obtain, 
\bea
\dot{\bm{\mathfrak{r}}}_{n \bm k} &=& D_{n \bm{k}}\biggl[ \tilde{\bm v}_{n\bm{k}} + \frac{e}{\hbar}({\bm v}_{n\bm{k}} \cdot {\bm{\Omega}}_{n\bm{k}})\bm{B} + \frac{\partial \bm{\cal R}_{n\bm{k}}^{B}}{\partial t} \biggr]~\label{EOM_r}, \\
\hbar \dot{\bm{k}} &=& D_{n\bm{k}}\left[ -e({\bm v}_{n\bm k} \times \bm{B}) \right] ~, \label{EOM_k}
\eea
where $D_{n\bm{k}} = [1+e/\hbar(\bm{B}\cdot \tilde{\bm{\Omega}}_{n\bm{k}})]^{-1}$ is the semiclassical phase space factor~\cite{Niu_05_prl}. The detailed derivation of EOM is presented in Appendix~\ref{Appendix: EOM}. In Eq.~(\ref{EOM_r}), the first term corresponds to the field-modified band velocity, the second term is the `chiral magnetic velocity' (CMV)~\cite{Tanizaki_17_prb} that gives rise to the chiral magnetic effect and the chiral anomaly~\cite{Sunit_23_prb} induced negative magneto-resistance in three-dimensional Weyl semimetals~\cite{Son_2013_PRB,Kamal_2019_PRB}. The third contribution is new and arises from the magnetic field perturbed wavefunction. It contains both spin and orbital components. While the spin part of the third term was discussed recently by Wang \emph{et al.}~\cite{Wang_2025_prl}, its orbital counterpart has not been reported before. Importantly, this velocity component, being the time derivative of the field-corrected Berry connection, contributes only in the presence of a dynamic magnetic field and vanishes for static driving. Additionally, in two-dimensional materials, the chiral velocity term vanishes due to the orthogonal configuration of electronic velocity and Berry curvature.

%%%%%%%%%%%%%%%%%%%%%%%%%%%%%%%%%%%%%%%%%%%%%%%%%%%%%%%%%%%%%%%%%%%%%%%

\begingroup 
\setlength{\tabcolsep}{8pt}
\begin{table*}[t!]
\caption{ The symmetry restrictions and physical origin of different contributions to the intrinsic gyrotropic magnetic current. The integrands of the current components are summed over all bands ($n$) and integrated over momentum ($\bm k$) space. This combined operation is denoted by $\int_{n\bm k}\equiv \sum_n \int d^d\bm{k}/ (2 \pi)^d$, where $d$ is the spatial dimension of the system. Here, $\cancel{\mathcal P}$ and $\cancel{\mathcal T}$ ($\mathcal P$, $\mathcal T$) indicate that spatial inversion and time-reversal symmetries are broken (preserved) in the system. The cross (\xmark) and the tick (\cmark) mark signify that the corresponding response tensor is symmetry forbidden and allowed, respectively. 
}
\begin{tabular}{c c c c c c c c}
\hline \hline 
\noalign{\vskip 6pt}
$\mathrm{Conductivity}$ & $\mathrm{Expression}$  & ${\mathcal{P} ,\cancel{\mathcal T}  }$ & $\cancel{\mathcal P},{\mathcal T}$ & $\cancel{\mathcal P},\cancel{\mathcal T}$ & $\mathcal{P}\mathcal{T}$ & $\mathrm{Physical \, Origin}$ \\
\noalign{\vskip 6pt}
\hline
\noalign{\vskip 6pt}

$\chi_{a;d}^{\mathrm{IG,Disp}}$ & $-ie \omega \int_{n\bm{k}} \tilde{\cal{G}}_{n}^{B;ad}(\omega) f_{n\bm k}$ & \xmark  & \xmark & \cmark & \cmark & $\bm{B}(t)$-induced Polarization Oscillation\\

\noalign{\vskip 6pt}

$\chi_{a;d}^{\mathrm{IG,CMV}}$ & $-\delta_{ad}(e^2/\hbar)\int_{n\bm k}({\bm v}_{n\bm{k}} \cdot {\bm{\Omega}}_{n\bm{k}})f_{n\bm k}$ & \xmark  & \cmark & \cmark & \xmark  & Chiral Magnetic Velocity\\

\noalign{\vskip 6pt}
\hline \hline
\end{tabular}
\label{Table_IGMC}
\end{table*}
\endgroup
%

%%%%%%%%%%%%%%%%%%%%%%%%%%%%%%%%%%%%%%%%%%%%%%%%%%%%%%%%%%%%%%%%%%

\section{Gyrotropic Magnetic Current} \label{Sec: GMC}

In the last section, we derived the particle velocity in the presence of an oscillating magnetic field. This velocity weighted by the nonequilibrium distribution function and the phase space correction factor, yields the charge current 
\be \label{j_def}
    \bm{j} = -e\int_{n\bm k} D_{n\bm k}^{-1}\dot{\bm{\mathfrak{r}}}_{n\bm{k}}\tilde{f}_{n\bm{k}}~.
\ee
For simplicity, we have defined $\int_{n\bm k}\equiv \sum_n \int d^d\bm{k}/ (2 \pi)^d$, with $d$ being the spatial dimension of the system. For a spatially homogeneous perturbation, the field-modified electronic distribution function $\tilde{f}_{n\bm k}$ is obtained by solving the Boltzmann equation under relaxation time approximation~\cite{Shengyuan_25_prl}, 
\bea
    \frac{\partial \tilde{f}_{n\bm k}}{\partial t} + \dot{\bm{k}}\cdot\nabla_{\bm{k}} \tilde{f}_{n\bm{k}} = -\frac{\tilde{f}_{n \bm{k}}- f_{n\bm k}^B}{\tau}~.\label{Boltzmann_eq}
\eea
Here, $f_{n\bm k}^B = \left(1+e^{\beta(\tilde{\varepsilon}_{n\bm{k}} - \mu)}\right)^{-1}$ is the Fermi-Dirac distribution function of the magnetic field perturbed band $\tilde{\varepsilon}_{n\bm{k}}$ with chemical potential $\mu$ and inverse temperature $\beta = 1/(k_BT)$. Due to periodic magnetic driving, the nonequilibrium correction part $\delta f_{n\bm k} = \tilde{f}_{n\bm k} - f_{n\bm k}^B$ would be in harmonics of the driving field. Solving Eq.~(\ref{Boltzmann_eq}) up to the linear order of the field, we obtain  $\delta f_{n\bm k}(t) = \delta f_{n\bm k}(\omega)e^{i\omega t} + \mathrm{c.c.}$, with
\bea
\delta f_{n\bm k}(\omega) &=& i\omega g^{\omega}_0 (\bm{{\cal M}}_{n\bm{k}} \cdot \bm{B}_0) \frac{\partial f_{n\bm k}}{\partial \varepsilon_{n\bm k}}~,
\eea
where $g^{\omega}_0 = [1/\tau+i \omega]^{-1}$, and $f_{n\bm k}$ corresponds to the equilibrium Fermi function without any magnetic field. Substitution of the field induced distribution function and the particle velocity from Eq.~(\ref{EOM_r}) into Eq.~(\ref{j_def}) and keeping terms only up to first order in the magnetic field yields the linear gyrotropic current. We express it as 
\bea \label{j_GMC}
j_a^{\mathrm{GMC}}(t) &=& \chi_{a;d}^{\mathrm{GMC}}(\omega)B_0^de^{i\omega t} + \mathrm{c.c.}~,
\eea
where $\{a,d\}$ refer to cartesian indices. The gyrotropic magnetic conductivity $\chi_{a;d}^{\mathrm{GMC}}(\omega)$ originates from three different physical processes: Fermi-surface oscillation, dynamic charge polarization, and chiral magnetic velocity. This allows us to express the gyromagnetic response as, 
\bea   \label{GMC}
\chi_{a;d}^{\mathrm{GMC}}(\omega) &=& \chi_{a;d}^{\mathrm{G,FO}} + \chi_{a;d}^{\mathrm{G,Disp}} + \chi_{a;d}^{\mathrm{G,CMV}}~. 
\eea
The Fermi surface oscillation term $\chi_{a;d}^{\mathrm{G,FO}}$ arises from the nonequilibrium distribution function and corresponds to the conventional gyrortropic current~\cite{Moore_2016_prl}
\bea
\chi_{a;d}^{\mathrm{G,FO}}(\omega) &=&  -ie\omega g^{\omega}_0 \int_{n\bm k} v_{n\bm{k}}^a {\cal M}_{n\bm{k}}^d \frac{\partial f_{n\bm{k}}}{\partial \varepsilon_{n\bm{k}}}~,\label{Fermi_oscillation}
\eea
where ${\bm{\cal M}}_{n\bm{k}} = ({\bm{\cal M}}_{n\bm{k}}^{L} + {\bm{\cal M}}_{n\bm{k}}^{S})$ denotes the momentum resolved total magnetic moment.

The displacement contribution, $\chi_{a;d}^{\mathrm{G,Disp}}$, originates from a magnetic field induced dynamic charge polarization,
\be
P_a(t) = -e \int_{n\bm k} {\mathcal R}_{n\bm k}^{B;a}(t) f_{n\bm k} = \mu_{a;d}(\omega)B_0^de^{i\omega t} + \mathrm{c.c.}
\ee
Here, $\mu_{a;d}(\omega)$ corresponds to the `magnetic polarizability',
\be
\mu_{a;d}(\omega) = -e \int_{n\bm k}
\tilde{\mathcal G}_{n\bm k}^{B;ad}(\omega) f_{n\bm k}~.
\ee
The temporal variation of the charge polarization generates a gyrotropic magnetic displacement current $\partial \bm{P}/\partial t$. The corresponding gyrotropic conductivity is given by 
\be
\chi_{a;d}^{\mathrm{G,Disp}}(\omega)
= -ie\omega \int_{n\bm k}
\tilde{\mathcal G}_{n\bm k}^{B;ad}(\omega) f_{n\bm k}~.\label{Displacement_contribution}
\ee
\noindent
Importantly, the gyrotropic conductivity is related to the `magnetic polarizability' through $\chi_{a;d}^{\mathrm{G,Disp}}(\omega) = i\omega \mu_{a;d}(\omega)$, providing a direct probe of the system’s polarization response to magnetic driving. This is one of the central results in this paper and constitutes a novel addition to the GMC.

The third term, $\chi_{a;d}^{\mathrm{G,CMV}}$, stems from the CMV
\bea
\chi_{a;d}^{\mathrm{G,CMV}}(\omega) &=& -\frac{e^2}{\hbar}\delta_{ad}\int_{n\bm k}({\bm v}_{n\bm{k}} \cdot {\bm{\Omega}}_{n\bm{k}})f_{n\bm k}~.\label{chiral_magnetic_velocity}
\eea
It drives a current strictly along the direction of the magnetic field. 

In addition to their physical origin, these components also differ in other aspects. Both $\chi_{a;d}^{\mathrm{G,FO}}$ and $\chi_{a;d}^{\mathrm{G,Disp}}$ are purely dynamic and vanish in the dc limit ($\omega \!\to\! 0$), whereas $\chi_{a;d}^{\mathrm{G,CMV}}$ can persist even for static fields. However, for static driving, the CMV-induced current or the chiral magnetic effect~\cite{Vilenkin_1980_prd,Fukushima_08_prb}, requires the system to be in an initial nonequilibrium state~\cite{Franz_13_prl,Yamamoto_15_PRD}. Even for a dynamical magnetic field, this current survives only in the presence of a chiral chemical imbalance~\cite{Kamal_20_prr}, typically observed in inversion-broken Weyl semimetal~\cite{Sumanta_15_prb,Anmol_18_prb}, and vanishes in other materials. Another differentiating aspect of the three contributions of GMC is that $\chi_{a;d}^{\mathrm{G,FO}}$ is a Fermi-surface response and finite only in metals or doped semiconductors. In contrast, the chiral magnetic part originates from the Fermi sea, and the displacement term contains both Fermi-surface and Fermi-sea contributions. Consequently, they enable finite gyrotropic responses even in insulating systems. In addition, the three conductivity components obey distinct symmetry constraints, which are discussed in Sec. \ref{Sec: Symmetry}.

We next focus on the intrinsic component of the GMC. The chiral-current contribution is purely intrinsic, while the conventional and displacement currents depend explicitly on the symmetric scattering time $\tau$. The $\tau$-dependence of the conventional term originates from $g^{\omega}_0$, whereas in the displacement current it arises from $g^{\omega}_{nm} = [1/\tau + i(\omega_{nm} + \omega)]^{-1}$. In the transport regime $(\omega\tau \ll 1)$, these factors approximate to $g^{\omega}_0 \approx \tau$ and $g^{\omega}_{nm} \approx [i(\omega_{nm}+ \omega)]^{-1}$. As a consequence, in this regime, the conventional GMC becomes purely extrinsic while the displacement contribution is an intrinsic response governed solely by the band-geometric MBCP tensor. We express the total intrinsic gyrotropic magnetic current (IGMC) as
\bea \label{j_IGMC}
j_a^{\mathrm{IGMC}}(t) &=& \chi_{a;d}^{\mathrm{IGMC}}(\omega)B_0^de^{i\omega t} + \mathrm{c.c.}~,
\eea
where the corresponding conductivity is given by
\bea  \label{chi_IGMC}
\chi_{a;d}^{\mathrm{IGMC}}(\omega) = \chi_{a;d}^{\mathrm{IG,Disp}} + \chi_{a;d}^{\mathrm{IG,CMV}}~.
\eea
The expressions of these contributions to the IGMC, together with their contrasting fundamental symmetry restrictions, are summarized in Table~\ref{Table_IGMC}. The detailed calculation of IGMC is presented in Appendix~\ref{Appendix: IGMC}.

%%%%%%%%%%    Symmetry of the Band-geometric Quantities     %%%%%%%%%%

\begingroup
\setlength{\tabcolsep}{6pt}
\begin{table}[b!]
\caption{The symmetry transformations of the band geometric quantities in momentum space under inversion ($\cal{P}$), time-reversal ($\cal T$) and combined parity-time reversal ($\cal{PT}$) operation.}
\begin{tabular}{c c c c}
\hline \hline 
\noalign{\vskip 2pt}
Quantity & ${\mathcal P}$ & ${\mathcal T}$ & $\mathcal{P}\mathcal{T}$  \\

\noalign{\vskip 2pt}
\hline 
\noalign{\vskip 6pt}

$\bm{\mathcal R}_{nm}(\bm k)$ & $-\bm{\mathcal R}_{nm}(-\bm k)$ & $\bm{\mathcal R}_{mn}(-\bm k)$ & $-\bm{\mathcal R}_{mn}(-\bm k)$ \\

\noalign{\vskip 6pt}

$\bm{v}_{nm}(\bm k)$ & $-\bm{v}_{nm}(-\bm k)$ & $-\bm{v}_{mn}(-\bm k)$ & $\bm{v}_{mn}(-\bm k)$ \\

\noalign{\vskip 6pt}

$\bm{\mathcal M}_{nm}(\bm k)$ & $\bm{\mathcal M}_{nm}(-\bm k)$ & $-\bm{\mathcal M}_{mn}(-\bm k)$ & $-\bm{\mathcal M}_{mn}(\bm k)$ \\

\noalign{\vskip 6pt}

\hline \hline
\end{tabular}
\label{table_1}
\end{table}
\endgroup
%

%%%%%%%%%%%%%%%%%%%%%%%%%%%%%%%%%%%%%%%%%%%%%%%%%%%%%%%%%%%%%%%%%%%%%%

\section{Symmetry Analysis}\label{Sec: Symmetry}

In this section, we analyze the symmetry properties of the intrinsic gyrotropic magnetic current. Having established the distinct microscopic origins of the individual IGMC channels, we first examine how the underlying band-geometric quantities transform under fundamental symmetry operations. This allows us to identify the channel-resolved symmetry restrictions governing the displacement, and chiral contributions to IGMC, and to distinguish which mechanisms can contribute in different symmetry classes of materials.

Under spatial inversion ($\mathcal P$), the electrical current reverses sign, $\bm j \rightarrow -\bm j$, while the magnetic field remains invariant, $\bm B \rightarrow \bm B$. Invariance of the linear response relation in Eq.~(\ref{j_IGMC}) therefore requires the gyrotropic magnetic conductivity tensor $\chi^{\mathrm{IGMC}}_{a;d}$ to change sign under $\mathcal P$. However, because conductivity is an intrinsic material property, it must remain invariant in a centrosymmetric crystal. As a consequence, a nonzero IGMC is symmetry allowed only in inversion-broken, i.e., noncentrosymmetric, systems.
Under time-reversal symmetry ($\mathcal T$), both the current and the magnetic field reverse sign, $\bm j \rightarrow -\bm j$ and $\bm B \rightarrow -\bm B$, leaving the overall IGMC response invariant. The gyrotropic magnetic current is therefore not prohibited by time-reversal symmetry and can arise in both magnetic and nonmagnetic materials.

While this establishes the general conditions for a nonvanishing IGMC, it does not specify which of the underlying microscopic mechanisms, namely, the magnetic displacement current and the chiral magnetic velocity, can contribute in a given material. To address this question, we examine the momentum-space parity of the conductivity integrands associated with the relevant band-geometric quantities. Using the transformation properties summarized in Table~\ref{table_1}, we find that in inversion-broken nonmagnetic crystals, the chiral channel is symmetry allowed, whereas the displacement contribution vanishes. In contrast, in $\mathcal{PT}$-symmetric systems, the chiral component is forbidden, leaving a purely displacement current driven intrinsic gyrotropic magnetic response.

%%%%%%%%%%%%%%%%%%%%%%%%%%%%%%%%%%%%%%%%%%%%%%%%%%%%%%%%%%%%%%%%%%%%%%
\begingroup
\setlength{\tabcolsep}{4.5 pt}
\begin{table}[t!]
\caption{ The symmetry restrictions of the magnetic gyrotropic response tensors. The cross (\xmark) and the tick (\cmark) mark signify that the corresponding response tensor is symmetry forbidden and allowed, respectively. Here, ${\cal M}_{a}$, ${\cal C}_{n}^a$ and ${\cal S}_{n}^a~({\cal C}_{n}^a{\cal M}_{a})$ represent  mirror, $n$-fold rotation, and $n$-fold roto-reflection symmetry operation along the $a$-direction for $a=\{x,y,z\}$, respectively. }
\begin{tabular}{c c c c c c c c c c c}
\hline \hline 
\noalign{\vskip 2pt}
IGMC & ${\cal M}_x$ & ${\cal M}_y$ & ${\cal M}_z$ & ${\cal C}_{n}^x$ & ${\cal C}_{n}^y$ & ${\cal C}_{n}^z$ & ${\cal S}_{4}^x$ & ${\cal S}_{4}^y$ & ${\cal S}_{4}^z$  & ${\cal S}_{6}^a$ \\
\noalign{\vskip 6pt}
\hline 

\noalign{\vskip 6pt} 

$\chi_{x;z}^{\mathrm{IGMC}}$ & \cmark & \xmark & \cmark  & \xmark & \cmark & \xmark & \xmark & \cmark  & \xmark & \xmark  \\

\noalign{\vskip 6pt}

$\chi_{y;z}^{\mathrm{IGMC}}$  & \xmark & \cmark & \cmark & \cmark & \xmark  & \xmark & \cmark & \xmark &\xmark & \xmark    \\

\noalign{\vskip 6pt}
\hline \hline
\end{tabular}
\label{table_nonmag_point_group}
\end{table}
\endgroup

%%%%%%%%%%%%%%%%%%%%%%%%%%%%%%%%%%%%%%%%%%%%%%%%%%%%%%%%%%%%%%%%%%%%%%%%%

Having established the fundamental symmetry properties of the individual IGMC channels, we now examine the additional constraints imposed by crystallographic point group operations. The transformation behavior of a linear response tensor under a point group operation is dictated by two attributes: (i) whether the tensor is polar or axial, and (ii) its time-reversal parity. From Eq. (\ref{j_IGMC}), which relates the polar current vector $\bm j$ to the axial magnetic field $\bm B$, it follows that the gyrotropic magnetic conductivity $\chi^{\mathrm{IGMC}}_{a;d}$ is a second-rank axial tensor\cite{newnham_symmetry}. 

Among the two contributions to IGMC, the chiral magnetic velocity term $\chi^{\mathrm{IG,CMV}}_{a;d}$ is even under time-reversal symmetry. In contrast, the displacement current contribution $\chi^{\mathrm{IG,Disp}}_{a;d}$ is odd under time-reversal. For a general crystallographic point group operation $\cal O$, these tensors transform as
\bea
\chi_{a';d'}^{\mathrm{IG,CMV}} &=& \det{\mathcal O}{\mathcal O}_{a'a}{\mathcal O}_{d'd}~\chi_{a;d}^{\mathrm{IG,CMV}}~, \\
\chi_{a';d'}^{\mathrm{IG,Disp}} &=& \eta_{\mathcal{T}}\det{\mathcal O}{\mathcal O}_{a'a}{\mathcal O}_{d'd}~\chi_{a;d}^{\mathrm{IG,Disp}}~.
\eea%
Here, $\eta_{\mathcal T}=+1$ for purely spatial operations ($\mathcal O=R$) and $\eta_{\mathcal T}=-1$ for magnetic operations involving time-reversal ($\mathcal O = R\mathcal T$).
Since the $\mathcal T$-even and $\mathcal T$-odd parts of IGMC response tensors transform identically under nonmagnetic point group operations, they obey the same symmetry constraints in this case. 
Accordingly, rather than listing separate symmetry tables for each contribution, we present the allowed components of the total IGMC tensor in Table~\ref{table_nonmag_point_group}. For magnetic point groups, however, the two sectors transform differently and must be treated separately. The corresponding symmetry conditions for $\chi^{\mathrm{IG,CMV}}_{a;d}$ and $\chi^{\mathrm{IG,Disp}}_{a;d}$ are summarized in Table~\ref{table_mag_point_group} of Appendix~\ref{Sec: mag_point_group}.

%%%%%%%%%%%%%%%%%%%%%%%%%%%%%%%%%%%%%%%%%%%%%%%%%%%%%%%%%%%%%%%%%%%%%%%%%

\section{IGMC in \texorpdfstring{$\ce{CuMnAs}$}{CuMnAs}} \label{Sec: CuMnAs}

%%%%%%%%%%%%%%%%%%%%%%%%%%%%%%%%%%%%%%%%%%%%%%%%%%%%%%%%%%%
\begin{figure*}[t!]
    \centering
    \includegraphics[width=.7\linewidth]{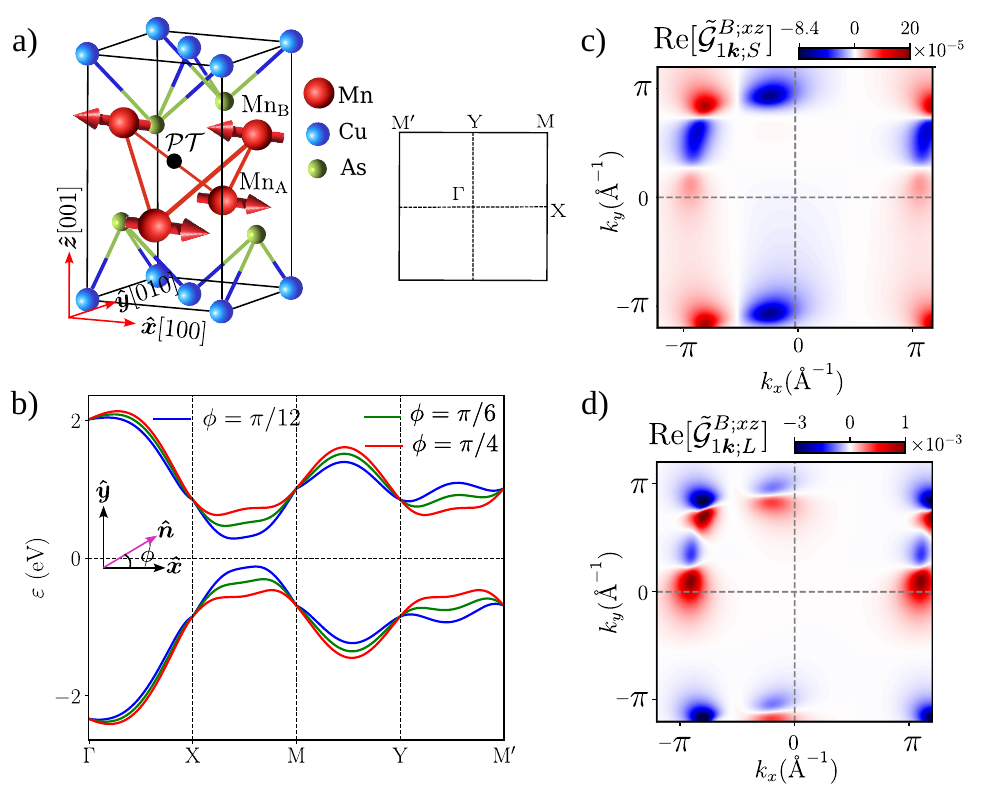}
    \caption{ (a) Crystallographic and magnetic structure of tetragonal $\ce{CuMnAs}$. The red arrows indicate the direction of magnetic moments. The projected Brillouin zone in the $x-y$ plane, along with high-symmetry points, is also shown. (b) The band dispersion for three different orientations of the N\'eel vector, $\phi = \{\pi/12, \pi/6,\pi/4\}$. Here, $\phi$ is the angle made by the N\'eel vector $(\hat{\bm n})$ with $\hat{\bm x}$ ([100]) axis. (c), (d) The momentum space distribution of \emph{spin}-MBCP ($\tilde{\cal{G}}_{1\bm{k};S}^{B;xz}$) and \emph{orbital}-MBCP ($\tilde{\cal{G}}_{1\bm{k};L}^{B;xz}$) for the first conduction band for N\'eel angle $\phi = \pi/4$.}
    \label{Fig_CuMnAs}
\end{figure*}
%%%%%%%%%%%%%%%%%%%%%%%%%%%%%%%%%%%%%%%%%%%%%%%%%%%%%%%%%%%

In tetragonal $\ce{CuMnAs}$~\cite{Smejkal_17_prl, Smejkal_23_npj}, the $\ce{Mn}$ atoms form two interpenetrating antiferromagnetic sublattices (denoted by \ce{Mn_A} and \ce{Mn_B}) as shown in Fig.~\ref{Fig_CuMnAs}(a). These magnetic atoms experience locally broken inversion symmetry, while being related to each other by the combined $\mathcal{PT}$ operation. The asymmetric local environment manifests via a sublattice-dependent anisotropic Rashba spin-orbit coupling (SOC). Due to the antiferromagnetic arrangement, the staggered magnetic moments are also subjected to a sublattice-contrasting exchange energy. Incorporating these energy contributions along with the intra- and inter-sublattice hopping, the minimal tight-binding Hamiltonian of $\ce{CuMnAs}$ layers can be written as~\cite{Watanabe_21_prx,Kamal_22_prl} 
\begin{align}
\mathcal{H}_0(\bm{k}) =
\begin{pmatrix}
\epsilon_0(\bm{k}) + \bm{h}_A(\bm{k})\!\cdot\!\bm{\sigma}  & V_{AB}(\bm{k}) \\
V_{AB}(\bm{k}) & \epsilon_0(\bm{k}) + \bm{h}_B(\bm{k})\!\cdot\!\bm{\sigma}
\end{pmatrix}.
\end{align}
Here, $\epsilon_0(\bm{k}) = -t(\cos k_x + \cos k_y)$ and $V_{AB}(\bm{k}) = -2\tilde{t}\cos(k_x/2)\cos(k_y/2)$ describes the hopping between the same and different sublattices, with $t=0.08$~eV and $\tilde{t}=1$~eV being the respective hopping amplitudes. The anisotropic SOC and exchange energy enter in the second term in the diagonal blocks $\bm{h}_A(\bm k) = -\bm{h}_B(\bm k) = \{ h_{\mathrm{AF}}^x - \alpha_{\mathrm{R}}\sin k_y, h_{\mathrm{AF}}^y + \alpha_{\mathrm{R}}\sin k_x, h_{\mathrm{AF}}^z\}$, where $\alpha_{\mathrm{R}} = 0.8$ eV is the SOC parameter. The N\'eel vector lies in the $xy$ plane, making an angle $\phi$ with the $x$ axis, $\hat{\bm n}=(\cos\phi,\sin\phi,0)$. Accordingly, we write the exchange field as $\bm{h}_{\mathrm{AF}} = \{h_{\mathrm{AF}}^x, h_{\mathrm{AF}}^y, h_{\mathrm{AF}}^z\} = J_n\hat{\bm n}$, with $J_n=0.85$~eV~\cite{Kamal_22_prl}. The energy eigenvalues of the above Hamiltonian are
\bea
\varepsilon_{\pm}(\bm k) &=& \epsilon_0(\bm{k}) \pm \sqrt{\abs{\bm{h}_A(\bm k)}^2 + V_{AB}(\bm{k})^2}~,
\eea
where $\pm$ correspond to the doubly degenerate conduction and valence bands, respectively. The resulting band gap,
\bea
\delta\varepsilon(\bm{k}) 
= 2\sqrt{|\bm{h}_A(\bm{k})|^2 + V_{AB}^2(\bm{k})}~,
\eea
is tunable through the orientation of the N\'eel vector. We present the band structure of $\ce{CuMnAs}$ in Fig.~\ref{Fig_CuMnAs}(b) for three different orientations of N\'eel vector.

Since the system preserves $\mathcal{PT}$ symmetry, $\chi_{a;d}^{\mathrm{IG,CMV}}$ is symmetry-forbidden, leaving the magnetic field–induced charge polarization oscillation as the sole source of IGMC. The momentum-space distributions of the underlying band geometric quantities, \emph{spin}-MBCP and \emph{orbital}-MBCP, are displayed in Figs. \ref{Fig_CuMnAs}(c) and \ref{Fig_CuMnAs}(d). The color scale of the plots indicates that the  \emph{orbital}-MBCP contribution is almost an order of magnitude larger than the \emph{spin}-MBCP contribution. 

\begin{figure*}[t!]
    \centering
    \includegraphics[width=.7\linewidth]{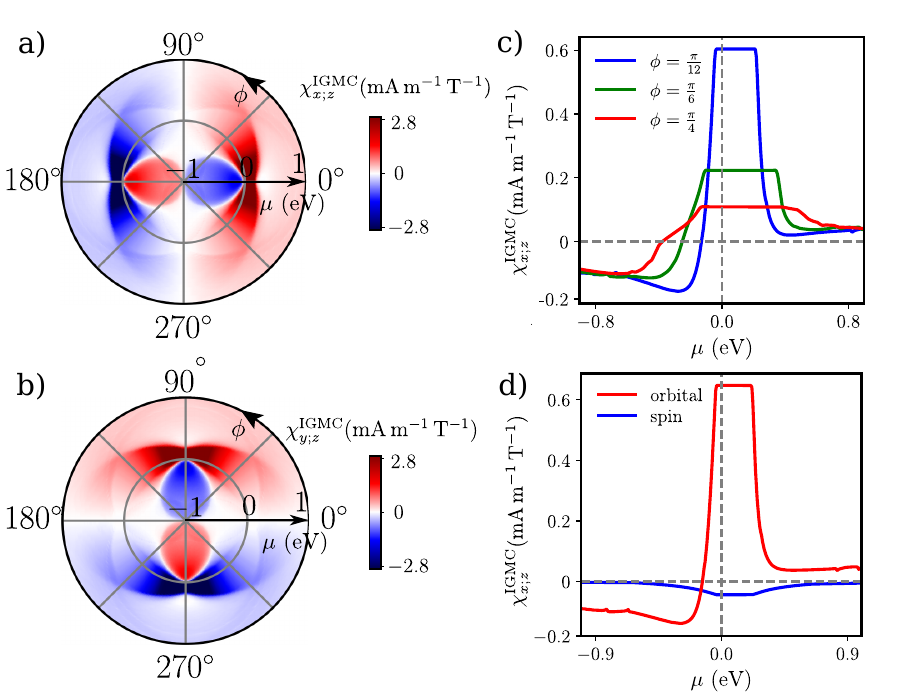}
    \caption{(a), (b): Intrinsic gyrotropic magnetic conductivities $\chi_{x;z}^{\mathrm{IGMC}}$ and $\chi_{y;z}^{\mathrm{IGMC}}$ as functions of the N\'eel vector angle $\phi$ and chemical potential $\mu$. The polar angle $\phi$ is measured from the $\hat{\bm{x}}$ axis, parallel to the crystallographic [100] direction. (c) Chemical potential dependence of $\chi_{x;z}^{\mathrm{IGMC}}$ for three representative N\'eel orientations. (d) \emph{orbital}- and \emph{Spin}-MBCP contributions to $\chi_{x;z}^{\mathrm{IGMC}}$ at $\phi = \pi/12$. For all plots, the magnetic driving frequency is assumed to be $\hbar \omega = 1 ~ \mu \text{eV}$. The orbital response dominates the spin contribution by almost an order of magnitude. The IGMC response in the band gap highlights that it is a Fermi sea response, and it measures the total `magnetic polarizability' of all the filled bands.} 
    \label{Fig_2}
\end{figure*}

For a magnetic field along the $\hat{\bm z}$ axis of the crystal, the IGMC tensors $\chi_{x;z}^{\mathrm{IGMC}}$ and $\chi_{y;z}^{\mathrm{IGMC}}$ are restricted by the microscopic symmetry of the material. In tetragonal $\ce{CuMnAs}$, the crystalline symmetry can be controlled by the N\'eel vector orientation. When the N\'eel vector aligns along $\hat{\bm{x}}$, the system posses $\mathcal{M}_x$ symmetry but breaks $\mathcal{M}_y$. Consequently, the system allows only $\chi^{\text{IGMC}}_{x;z}$ response but prohibits $\chi^{\text{IGMC}}_{y;z}$ (See Table \ref{table_nonmag_point_group}). Rotating the N\'eel vector to $\hat{\bm{y}}$ reverses the symmetry configuration: $\mathcal{M}_x$ is lost and $\mathcal{M}_y$ is preserved. This suppresses $\chi_{x;z}^{\mathrm{IGMC}}$, but allows $\chi_{y;z}^{\mathrm{IGMC}}$. A continuous rotation of the N\'eel vector between these two high-symmetry directions induces a smooth transfer of mirror symmetry, and the IGMC evolves accordingly: $\chi_{x;z}^{\mathrm{IGMC}}$ decreases while $\chi_{y;z}^{\mathrm{IGMC}}$ increases. This is in agreement with Figs. \ref{Fig_2}(a) and \ref{Fig_2}(b). The variation in the IGMC amplitude can also be attributed to the evolution of the \emph{spin}- and \emph{orbital}-MBCP tensors. A rotation of the N\'eel vector modifies the band gap, and because the MBCP tensors scale inversely with band gap, their magnitudes and the associated IGMC tensors change accordingly. From Fig.~\ref{Fig_2}(c), we note that for a fixed N\'eel orientation, $\chi_{x;z}^{\mathrm{IGMC}}(\omega)$ remains constant when the chemical potential lies within the band gap. This is due to the Fermi sea nature of IGMC.

%%%%%%%%%%%%%%%%%%%%%%%%%%%%%%%%%%%%%%%%%%%%%%%%%%%%%%%%%%%%%%%%

%%%%%%%%%%%%%%%%%%%%%%%%%%%%%%%%%%%%%%%%%%%%%%%%%%%%%%%%%%%%%%%

Notably, the IGMC conductivity in $\cal{PT}$-symmetric materials is a $\cal{T}$-odd response, and its sign reverses when the magnetization order, or the N\'eel vector, switches direction. This can be seen from Figs. \ref{Fig_2}(b). This symmetry-protected sign change makes IGMC a sensitive probe of the N\'eel vector orientation in $\ce{CuMnAs}$ and, more broadly, in other $\mathcal{PT}$-symmetric antiferromagnets.

We compare the \emph{spin}-MBCP and \emph{orbital}-MBCP contributions in Fig.~\ref{Fig_2}(d). Even though $\ce{CuMnAs}$ has a finite SOC, the spin contribution to IGMC arising from the $\tilde{\cal{G}}_{n;S}^{B;ad}$ is almost an order ($\sim 20$ times) of magnitude smaller than its orbital counterpart stemming from $\tilde{\cal{G}}_{n;L}^{B;ad}$. This highlights the significance of the orbital contribution for IGMC responses even in SOC materials.

To demonstrate the experimental feasibility of predicted results, we estimate the electric voltage arising from IGMC. We use the results from the 2D model for CuMnAs layers and convert the computed gyrotropic conductivity into an effective 3D conductivity by dividing it by interlayer spacing, $c = 6.32~ \text{\AA}$ for CuMnAs \cite{Wadley_13_natcomm}. The electrical voltage is given by $ V = \chi_{x;z}^{3D} B_0 \rho l $, with $\rho$ and $l$ are being the longitudinal resistivity and length of the sample, respectively and $ \chi^{3D}_{x;z} = (\chi_{x;z}^{\mathrm{IGMC}}/c) $. By taking a representative values of $ \rho = 50 ~\mu \Omega ~\text{cm}$  \cite{Wagenknecht_2020}, $l  = 100~\mu \text{m} $ and the maximum value of $\chi^{\text{IGMC}}_{x;z}$ from Fig.~(\ref{Fig_2}a), the electric voltage is estimated to be $V = 0.2 ~\text{mV}$ for a magnetic field strength $B_0 = 1 ~\text{T}$. This is well
within experimental reach.

%%%%%%%%%%%%%%%%%%%%%%%%%%%%%%%%%%%%%%%%%%%%%%%%%%%%%%%%%%%%%%%%%%%%%%%%%%%%

\section{Conclusion} \label{Sec: Conclusion}

To summarize, we have developed a microscopic theory of the gyrotropic magnetic current that unifies the orbital mechanism with the recently proposed spin-driven magnetic displacement contribution. Crucially, we identify the orbital component of the magnetic displacement current, which has not been explored earlier. 

Starting from a magnetic quantum kinetic equation with both spin-Zeeman and orbital coupling, we derived the finite-frequency semiclassical Lagrangian and the corresponding equations of motion. This naturally yields a magnetic field induced Berry connection correction, with both spin and orbital contributions. The orbital contribution includes an additional term that was overlooked in earlier semiclassical treatments. To quantify these effects, we introduce two band-geometric quantities: the spin and orbital magnetic Berry connection polarizabilities. When integrated over the Brillouin zone, the Berry-connection correction generates a magnetic field induced charge polarization. The temporal modulation of the charge polarization produces the magnetic displacement current. This displacement current can also have an intrinsic contribution, giving rise to IGMC. 

Our symmetry analysis reveals that this intrinsic gyrotropic magnetic current is finite in inversion-broken systems that preserve combined inversion and time-reversal symmetry, where the conventional gyrotropic conductivity vanishes. Guided by the symmetry analysis, we demonstrate the effect in tetragonal $\ce{CuMnAs}$. We find that the orbital contribution dominates over the spin channel by nearly an order of magnitude, establishing the orbital mechanism as the leading source of IGMC. Moreover, the IGMC response is tunable by the N\'eel vector orientation and, being time-reversal-odd, it reverses sign upon N\'eel vector reversal. This N\'eel vector dependent sign change identifies IGMC as a direct and experimentally accessible probe of antiferromagnetic order in $\ce{CuMnAs}$ and related bipartite antiferromagnets.

%%%%%%%%%%%%%%%%%%%%%%%%%%%%%%%%%%%%%%%%%%%%%%%%%%%%%%%%%%%%%%
\section{Acknowledgment}
We thank Sunit Das (IIT Kanpur, India) for many fruitful discussions. K.G. acknowledges the Ministry of Education, Government of India, for financial support through the Prime Minister’s Research Fellowship. S.S. is supported by the Indian Institute of Technology Kanpur. A.A. acknowledges funding from the Core Research Grant by ANRF (Sanction No. CRG/2023/007003), Department of Science and Technology, India.

%%%%%%%%%%%%%%%%%%%%%%%%%%%%%%%%%%%%%%%%%%%%%%%%%%%%%%%%%%%%%%

%      ##########         APPENDIX          ########       

%%%%%%%%%%%%%%%%%%%%%%%%%%%%%%%%%%%%%%%%%%%%%%%%%%%%%%%%%%%%%%

\onecolumngrid

\appendix

\section{Density matrix calculation in the presence of an oscillating magnetic field}\label{First_order_DM}

In this section, we study the dynamics of Bloch electrons in the presence of an oscillating magnetic field. In the absence of any external field, the distribution of Bloch electrons is characterized by the equilibrium Fermi--Dirac distribution function:
$f_{n} = [1 + e^{\beta(\varepsilon_{n \bm k} - \mu)}]^{-1},$
where $\varepsilon_{n \bm k}$ is the the $n^{\mathrm{th}}$ band energy eigenvalue, $\mu$ is the chemical potential, $\beta = 1/(k_BT)$ with $k_B$ is Boltzmann constant and $T$ is absolute temperature of the system. The external magnetic field drives the system into a nonequilibrium state, which gives rise to the response in experimental measurements. To describe this nonequilibrium state and to account for the effects of the external magnetic field, we use the nonequilibrium density matrix of the Bloch electrons, which is obtained by solving the quantum Liouville equation. In our study, we consider an oscillating magnetic field of frequency $\omega$, defined as:
\be \label{B}
\bm{B}(t) = \frac{1}{2}\bm{B}_0 ( e^{i\omega t} + e^{-i\omega t} ) ~.
\ee
This magnetic field interacts with an electron's magnetic moment in two different ways: (i) minimal coupling with orbital degree of freedom, and (ii) Zeeman coupling with spin degree of freedom. Considering these two types of couplings, the total magnetic interaction can be expressed as 
\bea
\hat{\mathcal{H}}^B &=& \hat{\mathcal{H}}^B_L + \hat{\mathcal{H}}^B_S = -\bm{B} \cdot \hat{\bm{\mathcal M}}^L - \bm{B} \cdot \hat{\bm{\mathcal M}}^S~,
\label{magnetic_interaction}
\eea
where, $\hat{\bm{\mathcal M}}^L = -(e/4)[\hat{\bm{r}}\times \hat{\bm v} - \hat{\bm v}\times \hat{\bm{r}}]$ is the orbital magnetic momentum operator, and $\hat{\bm{\mathcal M}}^S = -(\mathsf{g}_s\mu_B/2)\hat{\bm{\sigma}}$ is the spin magnetic moment operator. Here, $\mathsf{g}_s$ is the spin $g$-factor, $\mu_B$ is the Bohr magneton, and $\bm{\hat \sigma} = (\hat{\sigma}_x, \hat{\sigma}_y, \hat{\sigma}_z)$ corresponds to the Pauli spin matrices. In the presence of this magnetic field, the quantum Liouville equation can be written in the form~\cite{Dimi_PRB_17}:
\bea \label{Qkt_B_appendix}
\frac{\partial \hat{\rho}}{\partial t} + \frac{i}{\hbar} [\hat{\mathcal{H}}_0, \hat{\rho}] + \frac{1}{\tau}\hat{\rho} = {\mathcal D}_B \hat{\rho} - \frac{i\mathsf{g}_s \mu_B}{2\hbar} \bm{B} \cdot [\hat{\bm{\sigma}}, \hat{\rho}]~.
\eea
Here, $\hat{\mathcal{H}}_0$ denotes the unperturbed Bloch Hamiltonian, which satisfies 
$\hat{\mathcal{H}}_0 \ket{u_{n\bm{k}}} = \varepsilon_{n\bm{k}} \ket{u_{n\bm{k}}}$, 
with $\ket{u_{n\bm{k}}}$ being the cell-periodic part of the Bloch state. In Eq.~(\ref{Qkt_B_appendix}), $\tau$ is a phenomenological scattering time that characterizes the relaxation of the perturbed system. The first term on the right-hand side of Eq.~(\ref{Qkt_B_appendix}), $\mathcal{D}_B\hat{\rho}$, captures the interaction between the magnetic field and orbital moment. The matrix elements of this term, in the Bloch basis, are given by 
\bea \label{Db_rho_nm}
[\mathcal{D}_B\hat{\rho}]_{nm} &=& \frac{e}{2\hbar^2} \big\{ ( {\mathcal D}_{\bm k}\hat{\mathcal{H}}_0\times \bm{B})\cdot{\mathcal D}_{\bm k} \hat{\rho} \big\}_{nm} = \frac{e}{2\hbar^2} B^c ~ \epsilon_{cab} \big\{ {\mathcal D}_b\hat{\mathcal{H}}_0, {\mathcal D}_{a} \hat{\rho} \big\}_{nm}~.
\eea
In Eq.~(\ref{Db_rho_nm}), the notation $\{ \bm p ~ \cdot ~ \bm q \} = \bm p \cdot \bm q + \bm q \cdot \bm p$ (with $\bm p$ and $\bm q$ being vector operators) denotes a symmetrized operator product and $\mathcal{D}_a \hat{\mathcal{X}} \equiv  \mathcal{D} \hat{\mathcal{X}}/\mathcal{D}k_a$, where the action of covariant derivative $\mathcal{D}_{\bm k} \equiv \mathcal{D}/\mathcal{D}\bm{k}$ on a matrix operator $\hat{\mathcal{X}}$ is defined as follows:
\bea
[{\mathcal D}_{\bm k} \hat{\mathcal{X}}]_{nm} \equiv \left(\frac{{\mathcal D} \hat{\mathcal{X}}}{{\mathcal D}\bm{k}}\right)_{nm} = \partial_{\bm k}\mathcal{X}_{nm} - i[\bm{\hat{\mathcal{R}}},\hat{\mathcal{X}}]_{nm}~.
\eea
In the above equation, $\partial_{\bm k} \equiv \partial/\partial \bm k$-denotes simple derivative with respect to crystal momentum and $[\bm{\hat{\mathcal{R}}}]_{nm} \equiv \bm{\mathcal{R}}_{nm} = \bra{u_{n\bm k}}i\partial_{\bm k}\ket{u_{m\bm{k}}}$ is the band-resolved Berry connection. 

In the weak field limit, we can treat the magnetic interaction as a perturbation over the crystal Hamiltonian. In this situation, Eq.~(\ref{Qkt_B_appendix}) is solved perturbatively to obtain density matrix elements at different orders in magnetic field strength. The $N^{\text{th}}$ order correction to the density matrix,  $\hat{\rho}^{(B^N)} (~\sim\abs{\bm B}^N)$ can be expressed as, 
\bea
    (\partial_t + \mathcal{E}) \hat{\rho}^{(B^N)} &=& \mathcal{D}_{B}\hat{\rho}^{(B^{N-1})} - \frac{i\mathsf{g}_s \mu_B}{2\hbar} \bm{B} \cdot [\hat{\bm{\sigma}}, \hat{\rho}^{(B^{N-1})}]~, \label{EOM_rho_B}
\eea
where we have defined $\mathcal{E} = (\mathcal{P} + 1/\tau)$ with $\mathcal{P}\hat{\rho} = (i/\hbar)[\hat{\mathcal{H}_0},\hat {\rho}]$ for compactness of the equation.

\subsection*{Solution for first order density matrix equation}

The intrinsic gyrotropic magnetic current depends on the first-order density matrix elements. Inserting $N=1$ into Eq.~(\ref{EOM_rho_B}), we get the quantum Liouville equation for first order density matrix, $\hat{\rho}^B$, as follows:
\begin{align}
    \frac{\partial \hat{\rho}^B}{\partial t} +\frac{i}{\hbar}[\hat{\mathcal{H}_0},\hat {\rho}^B] + \frac{1}{\tau}\hat{\rho}^B = {\mathcal D}_B \hat{\rho}^{(0)} - \frac{i\mathsf{g}_s \mu_B}{2\hbar} \bm{B} \cdot [\hat{\bm{\sigma}}, \hat{\rho}^{(0)}]~.\label{QKT_B1}
\end{align}
Here, $\hat{\rho}^{(0)}$ denotes the density matrix of the Bloch electrons in the absence of a magnetic field, given by 
$\rho^{(0)}_{nm} = \delta_{nm} f_n$. In the steady state, the time dependence of the density matrix could only be carried in its' dynamic phase
\be
\rho_{nm}^B(t) =  {\rho}_{nm}^B(\omega) ~ e^{i\omega t} +  {\rho}_{nm}^B(-\omega) ~ e^{-i\omega t}~.
\ee
We substitute this \emph{ansatz} into Eq.~(\ref{QKT_B1}) to obtain the off-diagonal elements of the first-order density matrix, which are
\begin{align}
\rho_{nm}^B (\omega) &= \frac{1}{\frac{1}{\tau}+i(\omega_{nm}+\omega)} \frac{e}{4\hbar^2} \big\{ ( {\mathcal D}_{\bm k}\hat{\mathcal{H}}_0\times \bm{B}_0 ) \cdot{\mathcal D}_{\bm k} \hat{\rho}^{(0)} \big\}_{nm} - \frac{1}{\frac{1}{\tau}+i(\omega_{nm}+\omega)} \frac{i\mathsf{g}_s \mu_B}{4\hbar} \bm{B}_0 \cdot [\hat{\bm{\sigma}}, \hat{\rho}^{(0)}]_{nm}      \nn \\
&= \frac{e}{4\hbar^2}~g_{nm}^{\omega} \epsilon_{cab} ~B_0^c~ \big\{ {\mathcal D}_b\hat{\mathcal{H}}_0, {\mathcal D}_{a} \hat{\rho}^{(0)} \big\}_{nm} +  \frac{i\mathsf{g}_s \mu_B}{4\hbar}~ g_{nm}^{\omega}(\bm{B}_0 \cdot \bm{\sigma}_{nm}) f_{nm} \nn \\
&= \rho_{nm;L}^B (\omega) + \rho_{nm;S}^B (\omega)~,\label{rho_B_OffDiag}
\end{align}
where $g_{nm}^{\omega} = [1/\tau+i(\omega_{nm} + \omega)]^{-1}$ and $f_{nm} = f_n - f_m$. The matrix elements of the covariant derivative of $\mathcal{H}_0$ and $\hat{\rho}^{(0)}$ are given by
\bea
[{\mathcal D}_b \hat{\mathcal{H}}_0]_{nm} &=& 
\hbar\left[ \delta_{nm}v_{n}^b + i\omega_{nm}\mathcal{R}_{nm}^b \right]
= \hbar\left[ \delta_{nm}v_{n}^b + v_{nm}^b \right], \label{D_H0} \\
\left[ \mathcal{D}_a\hat{\rho}_0 \right]_{nm} &=&
\left[ \delta_{nm}(\partial_a f_n) + i f_{nm}\mathcal{R}_{nm}^a \right]~,
\label{D_rho_0}
\eea
where, $v_{nm}^b = i\omega_{nm}\mathcal{R}_{nm}^b$ is the off-diagonal velocity matrix element. The first part of Eq.~(\ref{rho_B_OffDiag}), $\rho_{nm;L}^B (\omega)$, corresponds to the orbital contribution to the off-diagonal density matrix and the second part, $\rho_{nm;S}^B (\omega)$,  depicts the spin contribution. Notably, the spin part depends on the off-diagonal spin matrix element. Therefore, this contribution depends on the spin-orbit coupling (SOC) strength of the material and would be negligible in materials with low SOC. While the orbital part is not restricted to any such condition, it would be prevalent in all materials. Substitution of the covariant derivatives from Eqs. (\ref{D_H0}) and (\ref{D_rho_0}) into the orbital part yields
\begin{align}
\rho_{nm;L}^B (\omega)
 &= \frac{e}{4\hbar^2} g_{nm}^{\omega} \epsilon_{cab} B_0^c  
    \left\{ {\mathcal D}_b\hat{\mathcal{H}}_0, {\mathcal D}_{a} \hat{\rho}^{(0)} \right\}_{nm}  \nonumber\\[4pt]
&= \frac{e}{4\hbar^2} g_{nm}^{\omega} \epsilon_{cab} B_0^c  
   \sum_p \Bigl( [{\mathcal D}_b \hat{\mathcal{H}}_0]_{np} [{\mathcal D}_{a} \hat{\rho}^{(0)}]_{pm}  
   + [{\mathcal D}_{a} \hat{\rho}^{(0)}]_{np} [{\mathcal D}_b\hat{\mathcal{H}}_0]_{pm} \Bigr)  \nonumber\\[4pt]
&= \frac{e}{4\hbar^2} g_{nm}^{\omega} \epsilon_{cab} B_0^c  
   \sum_p \Bigl( \hbar[\delta_{np}v_{n}^b + v_{np}^b]
        [\delta_{pm}(\partial_a f_m) + i f_{pm}\mathcal{R}_{pm}^a] + [\delta_{np}(\partial_a f_n) + i f_{np}\mathcal{R}_{np}^a]
       \hbar[\delta_{pm}v_{m}^b + v_{pm}^b] \Bigr)  \nonumber\\[4pt]
&= \frac{e}{4\hbar} g_{nm}^{\omega} \bm{B}_0 \!\cdot\!
   \left[\partial_{\bm k}(f_n + f_m) \!\times\! \bm{v}_{nm}\right] - \frac{i e}{4\hbar} g_{nm}^{\omega} \bm{B}_0 \!\cdot\!
   \sum_{p \neq m} (\bm{v}_{np} + \delta_{np}\bm{v}_m) \!\times\! \bm{\mathcal R}_{pm} f_{pm}  \nonumber\\
&\quad
   + \frac{i e}{4\hbar} g_{nm}^{\omega} \bm{B}_0 \!\cdot\!
   \sum_{p \neq n} \bm{\mathcal R}_{np} \!\times\!
   (\bm{v}_{pm} + \delta_{pm}\bm{v}_n) f_{np}~.
\label{rho_nm_B_v2_appendix}
\end{align}
From Eq.~(\ref{rho_B_OffDiag}), the spin part of the off-diagonal density matrix is given by
\bea  \label{rho_B_S}
\rho_{nm;S}^B (\omega) &=& \frac{i\mathsf{g}_s \mu_B}{4\hbar} g_{nm}^{\omega}(\bm{B}_0 \cdot \bm{\sigma}_{nm})f_{nm}~.
\eea
%

%%%%%%%%%%%%%%%%%%%%%%%%%%%%%%%%%%%%%%%%%%%%%%%%%%%%%%%%%%%%%%%%%%%%%%%%%

\section{Derivation of semiclassical Lagrangian from density matrix framework}\label{Appendix B: Lagrangian}

The semiclassical transport theory presents a simplified understanding of intraband dynamics of charge carriers under small perturbations. While the density matrix framework extends the understanding by including the interband dynamics. The motivation of this section is to unify these two different pictures by deriving the equation of motion (EOM) from the density matrix formalism in the semiclassical limit. Semiclassical description is valid only in the regime where (i) the spatial variations of the external fields are slow over the wavepacket extension, and (ii) the temporal variation doesn't induce any real interband transition. We respect the first criterion by assuming a standard semiclassical perturbation in the density matrix picture, and maintain the second by reducing terms into an effective single-band form. The basic recipes that we follow are: (i) to derive the Lagrangian using density matrix formalism and 
reduce it into a single-band expression, and then (ii) use the Euler-Lagrangian equations to derive the EOM. The Lagrangian operator for a system can be defined as follows:
\bea
\hat{\mathcal{L}} = i\hbar\frac{\partial}{\partial t} - \hat{\mathcal{H}}~.
\eea
In the density matrix picture, the Lagrangian of the system can be calculated following a trace of the combined Lagrangian and density matrix operator over the Bloch states: 
\bea \label{Lag_initial}
\mathcal{L}_{\bm k} &=& \Tr\{\hat{\mathcal{L}}\hat{\rho}\}~\nn \\
&=& \sum_n \bra{u_{n\bm k}} \left( i\hbar\frac{\partial}{\partial t} - \hat{\mathcal{H}} \right) \hat{\rho} \ket{u_{n\bm k}} \nn \\
&=& \sum_n \bra{u_{n\bm k}} \left( i\hbar\frac{\partial \hat{\rho}}{\partial t} + i\hbar \hat{\rho} \frac{\partial}{\partial t} - \hat{\mathcal{H}}\hat{\rho} \right) \ket{u_{n\bm k}} \nn \\
&=&  i\hbar \sum_n \bra{u_{n\bm k}} \frac{\partial \hat{\rho}}{\partial t} \ket{u_{n\bm k}} + i\hbar \sum_n \bra{u_{n\bm k}}\hat{\rho} \ket{\frac{\partial u_{n\bm k}}{\partial t}} - \sum_n \bra{u_{n\bm k}} \hat{\mathcal{H}}\hat{\rho} \ket{u_{n\bm k}} \nn \\
&=& i\hbar \sum_n \frac{\partial {\rho}_{nn}}{\partial t} + i\hbar \sum_{np} \bra{u_{n\bm k}}\hat{\rho}\ket{u_{p\bm k}}\bra{u_{p\bm k}} \ket{\frac{\partial u_{n\bm k}}{\partial t}} - \sum_n\bra{u_{n\bm k}} \hat{\mathcal{H}}\hat{\rho} \ket{u_{n\bm k}} \nn \\
&=& i\hbar \sum_n \frac{d {\rho}_{nn}}{dt}  -  i\hbar \sum_n \dot{\bm{k}} \cdot \frac{\partial {\rho}_{nn}}{\partial \bm{k}} + i\hbar \sum_{np} \rho_{np} \bra{u_{p\bm k}} \ket{\frac{\partial u_{n\bm k}}{\partial t}} - \sum_n\bra{u_{n\bm k}} \hat{\mathcal{H}}\hat{\rho} \ket{u_{n\bm k}} \nn \\
&=& - i\hbar \sum_n \dot{\bm{k}} \cdot \frac{\partial {\rho}_{nn}}{\partial \bm{k}} + i\hbar \sum_{np} \rho_{np} \bra{u_{p\bm k}} \ket{\frac{\partial u_{n\bm k}}{\partial t}} - \sum_n\bra{u_{n\bm k}} \hat{\mathcal{H}}\hat{\rho} \ket{u_{n\bm k}}~. 
\eea 
In the above expression, we have dropped the unnecessary total time derivative of the nonequilibrium density matrix ${  \rho}_{nn}$, since that would not affect the result of the Euler-Lagrangian equations. The first term involving the crystal momentum derivative of the density matrix can be expressed in terms of the position operator expectation value. The average electronic position $\bm{\mathfrak{r}}$ within an unit cell can be defined as follows,  
\bea \label{r}
\bm{\mathfrak{r}} = \expval{\hat{\bm{r}}} = \Tr \{\hat{\bm{r}}\hat{\rho}\} &=& \sum_{np}\bm{r}_{np}{\rho}_{pn} ~
= ~\sum_{np}(i\delta_{np}\partial_{\bm k} +  \bm{\mathcal R}_{np} ){\rho}_{pn}~,\nn \\
\Rightarrow{}\sum_n\bm{\mathfrak{r}}_n \rho_{nn} &=& \sum_n \left[i\frac{\partial {\rho}_{nn}}{\partial \bm{k}} + \sum_{p} {\rho}_{np}\bm{\mathcal R}_{pn} \right] \label{r_diag} \nn\\
\Rightarrow{} i\frac{\partial {\rho}_{nn}}{\partial \bm{k}} &=& \bm{\mathfrak{r}}_n \rho_{nn} - \sum_{p}{\rho}_{np}\bm{\mathcal R}_{pn}~,
\eea
where we have defined $\bm{\mathfrak{r}} = \sum_n \bm{\mathfrak{r}}_n \rho_{nn}$ with $\bm{\mathfrak{r}}_n$ being the effective single band expression of $\bm{\mathfrak{r}}$. Substituting the result from Eq.~(\ref{r}) into Eq.~(\ref{Lag_initial}) leads to the final
expression of Lagrangian, 
\bea 
\mathcal{L}_{\bm k} &=& -\sum_n \hbar\dot{\bm{k}}\cdot \left( \bm{\mathfrak{r}}_n \rho_{nn} - \sum_{p}{ \rho}_{np}\bm{\mathcal R}_{pn} \right) + i\hbar \sum_{np} \rho_{np} \bra{u_{p\bm k}} \ket{\frac{\partial u_{n\bm k}}{\partial t}} - \sum_n\bra{u_{n\bm k}} \hat{\mathcal{H}}\hat{\rho} \ket{u_{n\bm k}}~. \label{Lag_final}
\eea
The Bloch states in the presence of a magnetic field take the form $e^{i\bm{k}\cdot\bm{r}} \ket{u_n(\bm{k} + e\bm{A}(\bm{\mathfrak{r}})/\hbar)}$, where $\bm{A} = \frac{1}{2} (\bm{B} \times \bm{r})$ is the magnetic vector potential in symmetric gauge. In the presence of a magnetic field, $\bm k$ becomes gauge-dependent, and thereby we replace $\bm k$ in Eq.~(\ref{Lag_final}) with the gauge-invariant crystal momentum $\bm{\kappa} = \bm{k} + (e/\hbar)\bm{A}(\bm{\mathfrak{r}})$. Here, we denote the charge of an electron by $-e$, with $e>0$. 

Now, we evaluate each of the three terms in Eq.~(\ref{Lag_final}) separately. For the first term, we replace $\bm{k}$ with $\bm{\kappa}$ and use Eq.~(\ref{r}) to obtain

\bea  \label{Lag_B_term1}
 \sum_n \hbar\dot{\bm{k}} \cdot \left(- i\frac{\partial {\rho}_{nn}}{\partial \bm{k}}\right) &=& -\sum_n \left[ \hbar\dot{\bm{\kappa}} - \frac{e}{2}\bm{B}(t)\times \dot{\bm{\mathfrak r}}_n \right]\cdot\left[\bm{\mathfrak{r}}_n \rho_{nn} - \sum_{p}{\rho}_{np}\bm{\mathcal R}_{pn} \right]~.
\eea
We calculate the second term in Eq.~(\ref{Lag_final}) using the completeness of the Bloch basis and the partial time derivative of the Bloch state:
\bea  \label{Lag_B_term2}
i\hbar \sum_{np} \rho_{np} \bra{u_{p\bm{\kappa}}}\ket{\frac{\partial u_{n\bm{\kappa}}}{\partial t}} &=& i\sum_n \rho_{np} \left(\hbar\frac{\partial \bm{\kappa}}{\partial t}\right) \cdot \bra{u_{p\bm{\kappa}}}\ket{\frac{\partial u_{n\bm{\kappa}}}{\partial {\bm{\kappa}}}} \nn \\
&=& \frac{e}{2}(\bm{B}\times \dot{\bm{\mathfrak r}}_n) \cdot \sum_p {\rho}_{np}\bm{\mathcal R}_{pn}~.
\eea
The third term of Eq.~(\ref{Lag_final}) corresponds to the magnetic field induced total energy: 
\bea \label{E_B}
\sum_n\bra{u_{n\bm \kappa}} \hat{\mathcal{H}}\hat{\rho} \ket{u_{n\bm \kappa}} = \sum_n \tilde{\varepsilon}_{n\bm{\kappa}} \rho_{nn}~.
\eea
Here, $\tilde{\varepsilon}_{n\bm{\kappa}}$ denotes the field-corrected band energy. Substituting the results from Eqs. (\ref{Lag_B_term1}), (\ref{Lag_B_term2}) and (\ref{E_B}) into Eq.~(\ref{Lag_final}) yields the total Lagrangian, 
\begin{align}
\mathcal{L}_{\bm \kappa} &= -\sum_n \hbar\dot{\bm{\kappa}} \cdot \big[\bm{\mathfrak{r}}_n \rho_{nn} - \big({\rho}_{nn}\bm{\mathcal R}_{n\bm{k}} + \sum_{p \neq n}{\rho}_{np}(t)\bm{\mathcal R}_{pn} \big)\big]  + \frac{e}{2} \sum_n [ \bm{B}(t) \times \dot{\bm{\mathfrak r}}_n ] \cdot \bm{\mathfrak{r}}_n \rho_{nn} - \sum_n \tilde{\varepsilon}_{n\bm{\kappa}}\rho_{nn}~. \label{Lag_B_v1}
\end{align}
To extract the band-resolved Lagrangian from the total Lagrangian, we need to recast the second term inside the parentheses as an effective single-band quantity. Interestingly, this term yields a magnetic field induced  Berry connection, as derived in Eq.~(\ref{Berry_correction_total}). Using the expression from Eq.~(\ref{Berry_correction_total}), the total Lagrangian can be written as 
\begin{align}
\mathcal{L}_{\bm \kappa} &= -\sum_n \hbar\dot{\bm{\kappa}} \cdot \left[\bm{\mathfrak{r}}_n \rho_{nn} - \left( \bm{\mathcal R}_{n\bm{k}}{\rho}_{nn} + \bm{\mathcal{R}}_{n\bm{k}}^{B}(t)f_n\right)\right]  + \frac{e}{2} \sum_n [ \bm{B}(t) \times \dot{\bm{\mathfrak r}}_n ] \cdot \bm{\mathfrak{r}}_n \rho_{nn} - \sum_n \tilde{\varepsilon}_{n\bm{\kappa}}\rho_{nn}~.\label{LB_total}
\end{align}
For simplicity, from now onwards, we represent the gauge-invariant crystal momentum $\bm{\kappa}$ by $\bm{k}$. Recently, the unification~\cite{Jian_Wang_2024_PRB,Mehraeen_2024_PRB} of quantum kinetic framework and semiclassical transport theory for electric driving has been demonstrated from the perspective of charge current: $\bm{J} = \Tr(\hat{\bm v} \hat{\rho}) = \sum_n (\hat{\bm v} \hat{\rho})_{nn} = \sum_n \dot{\bm{\mathfrak{r}}}_n g_n$, where $\dot{\bm{\mathfrak{r}}}_n$ is the wavepacket velocity and $g_n$ corresponds to the nonequilibrium Boltzmann distribution function. Similarly, if the semiclassical band-momentum resolved Lagrangian be $\mathcal{L}_{n\bm k}^{\rm S}$, momentum resolved total Lagrangian in presence of a magnetic field would be $\mathcal{L}_{\bm k} = \sum_n \mathcal{L}_{n\bm k}^{\rm S}{D}_{n\bm k}^{-1}f_{n\bm k}$, where $D_{n\bm{k}}^{-1} = [1+e/\hbar(\bm{B} \cdot \bm{\Omega}_{n\bm{k}})]$ is the magnetic field corrected phase factor. The magnetic field induced diagonal density matrix can be calculated to be $\rho_{nn} =  [1+e/\hbar(\bm{B} \cdot \bm{\Omega}_{n\bm{k}})]f_{nk} = D_{n\bm{k}}^{-1} f_{nk}$~\cite{Dimi_PRB_17}. Thereby, the correct quantity to compare with $\mathcal{L}_{n\bm k}^{\rm S}$ would be $\mathcal{L}_{n\bm k}$, defined in the following way, 
\bea \label{LB_sum_Ln}
\mathcal{L}_{\bm k} = \sum_n \mathcal{L}_{n\bm k}\rho_{nn}~.
\eea
Now, comparing Eqs. (\ref{LB_total}) and (\ref{LB_sum_Ln}), and collecting terms up to second order in the magnetic field strength, we obtain the band-resolved Lagrangian to be
\begin{equation} \label{Lag_B_t}
\mathcal{L}_{n\bm k} = - \hbar\dot{\bm{k}} \cdot \left[\bm{\mathfrak{r}}_n - \left( \bm{\mathcal R}_{n\bm{k}} + \bm{\mathcal{R}}_{n\bm{k}}^{B}(t)\right)\right] + \frac{e}{2} [ \bm{B}(t) \times \dot{\bm{\mathfrak r}}_n ] \cdot \bm{\mathfrak{r}}_n - \tilde{\varepsilon}_{n\bm{k}}~. 
\end{equation}
To our knowledge, this is the first work to capture the effect of the time dependence of a magnetic field via minimal coupling interaction. In the dc limit, this result matches the existing semiclassical theory~\cite{Niu_prl_14,Gao_frontiers_19}, except that we obtain an extra term in the magnetic field induced Berry connection. The expression of the \textit{a.c} magnetic field corrected Berry connection, along with its zero-frequency result, is derived below.

%%%%%%%%%%%%%%%%%%%%%%%%%%%%%%%%%%%%%%%%%%%%

\subsection*{Magnetic field induced Berry connection correction}

Here, we aim to derive the density matrix formulation of the magnetic field induced Berry connection correction. For this, we plug in the magnetic field induced off-diagonal density matrix elements from Eq.~(\ref{rho_B_OffDiag}) into $\sum_{n,m \neq n} {\rho}_{nm}{\mathcal R}_{mn}^a$, and express the resultant expression into an effective single band term. Because the density matrix accounts for both minimal coupling and Zeeman interaction, the resulting Berry connection correction contains contributions from both orbital and spin magnetic moments, 
\bea \label{rho_R}
\sum_{n,m\neq n} {\rho}_{nm}^B(t){\mathcal R}_{mn}^a &=& \left[ \sum_{n,m\neq n}{\rho}_{nm;L}^B(\omega){\mathcal R}_{mn}^ae^{i\omega t} +  \sum_{n,m\neq n}{\rho}_{nm;S}^B(\omega){\mathcal R}_{mn}^a e^{i\omega t} \right] + (\omega \rightarrow -\omega)~.
\eea
As shown below, the first term in Eq.~(\ref{rho_R}) gives rise to the orbital contribution, while the second term yields the spin contribution to the Berry connection correction. For convenience, we calculate these two contributions separately.

%%%%%%%%%%%%%%%%%%%%%%%%%%%%%%%%%%%

\subsubsection{Orbital part of the Berry connection correction}
The orbital part of the magnetic field driven Berry connection correction can be obtained by evaluating the term
\bea  \label{Berry_L_int}
\sum_{n,m\neq n} {\rho}_{nm;L}^B(t){\mathcal R}_{mn}^a &=& \sum_{n,m\neq n}{\rho}_{nm;L}^B(\omega){\mathcal R}_{mn}^a e^{i\omega t} + (\omega \rightarrow - \omega)~.
\eea
Here, we have 
\bea
 \sum_{n,m\neq n} \rho_{nm;L}^B (\omega)  ~ {\mathcal R}_{mn}^a e^{i\omega t} &=&  e^{i\omega t}\frac{e}{ 4 \hbar} \sum_{n,m\neq n} g_{nm}^{\omega} \bm{B}_0 \cdot \left[ \{\partial_{\bm k}(f_n+f_m)\} \times \bm{v}_{nm}\right]{\mathcal R}_{mn}^a   \nn \\
&&  -~ e^{i\omega t} \frac{ie}{4\hbar} \sum_{n,m\neq n} g_{nm}^{\omega} \bm{B}_0 \cdot \sum_{p \neq m}[(\bm{v}_{np} + \delta_{np}\bm{v}_m)\times \bm{\mathcal R}_{pm}]{\mathcal R}_{mn}^a f_{pm}  \nn \\
&& +~ e^{i\omega t}\frac{ie}{4\hbar}   \sum_{n,m\neq n} g_{nm}^{\omega} \bm{B}_0 \cdot \sum_{p \neq n} [\bm{\mathcal R}_{np} \times (\bm{v}_{pm} + \delta_{pm}\bm{v}_n)]{\mathcal R}_{mn}^a f_{np}~. 
\eea
Following the invariance of dummy indices, we exchange $m\leftrightarrow n$ in the first term. For the second and third terms, we unpack the population difference and exchange indices $p\leftrightarrow n$ into the terms associated with $f_p$, thereby recasting the entire expression as an effective single-band summation
\bea  \label{Berry_L_mid}
\sum_{n,m\neq n} \rho_{nm;L}^B (\omega)  ~ {\mathcal R}_{mn}^a e^{i\omega t} 
&=& B_0^d e^{i\omega t}\frac{e}{4\hbar} \left[ \sum_{n,m\neq n} g_{nm}^{\omega} \left[ \epsilon_{dbc}(\partial_b f_n) v_{nm}^c\right]{\mathcal R}_{mn}^a + \sum_{n,m\neq n} (g_{nm}^{-\omega})^* \left[ \epsilon_{dbc}(\partial_b f_n) v_{mn}^c\right]{\mathcal R}_{nm}^a \right] \nn \\
&& - B_0^d e^{i\omega t} \frac{i}{2\hbar}  \sum_{n,m\neq n} g_{mn}^{\omega} \left[ -\frac{e}{2}\sum_{p \neq n}(\bm{v}_{mp} + \delta_{mp}\bm{v}_n)\times \bm{\mathcal R}_{pn}\right]^d{\mathcal R}_{nm}^a (f_n - f_p) \nn \\
&& + B_0^d e^{i\omega t} \frac{i}{2\hbar} \sum_{n,m\neq n} (g_{mn}^{-\omega})^* \left[- \frac{e}{2}\sum_{p \neq n} (\bm{v}_{pm} + \delta_{pm}\bm{v}_n) \times \bm{\mathcal R}_{np}\right]^d{\mathcal R}_{mn}^a (f_n - f_p)~.\nn \\
\eea 
Simplifying a bit, we can express the above equation as, 
\bea 
\sum_{n,m\neq n} \rho_{nm;L}^B (\omega)  ~ {\mathcal R}_{mn}^a e^{i\omega t} 
&=& \sum_n f_n \left[ \left( {\cal{G}}_{n\bm k;L}^{B;ad}(\omega) + [{\cal{G}}_{n\bm k;L}^{B;ad}(-\omega)]^*\right) B_0^d e^{i\omega t} \right]~,
\eea
where we have defined, 
\bea
{\cal{G}}_{n\bm k;L}^{B;ad}(\omega) &=& -\frac{e}{4\hbar}\sum_{m\neq n} \epsilon_{dbc} \partial_b (g_{nm}^{\omega} v_{nm}^c{\mathcal R}_{mn}^a) -\frac{i}{2\hbar} \sum_{m\neq n} g_{mn}^{\omega} \cal{M}_{mn}^{L;d}{\mathcal R}_{nm}^a + \frac{i}{2\hbar} \sum_{m\neq p} \sum_{p \neq n} g_{mp}^{\omega}\cal{M}_{mnp}^{L;d}{\mathcal R}_{pm}^a~.
\eea
Here, $\bm{\cal{M}}_{mnp}^L = -(e/2)(\bm{v}_{mn} + \delta_{mn}\bm{v}_p)\times \bm{\cal{R}}_{np}$, and $\bm{\cal{M}}_{mn}^L = \sum_{p\neq n}\bm{\cal{M}}_{mpn}^L$ is the matrix element of the electronic orbital magnetic moment operator. Substitution of the result from Eq.~(\ref{Berry_L_mid}) into Eq.~(\ref{Berry_L_int}) yields, 
\bea  \label{Berry_L_mid2}
\sum_{n,m\neq n} {\rho}_{nm;L}^B(t){\mathcal R}_{mn}^a &=& \sum_n f_n \left[ \left( {\cal{G}}_{n\bm k;L}^{B;ad}(\omega) + [{\cal{G}}_{n\bm k;L}^{B;ad}(-\omega)]^*\right) B_0^d e^{i\omega t} \right] + (\omega \rightarrow - \omega)~, \nn \\
&=& \sum_n f_n \left[ \left( {\cal{G}}_{n\bm k;L}^{B;ad}(\omega) + [{\cal{G}}_{n\bm k;L}^{B;ad}(-\omega)]^*\right) B_0^d e^{i\omega t} + \mathrm{c.c.} \right] \nn \\
&=& \sum_n f_n \left[ \tilde{\cal{G}}_{n\bm k;L}^{B;ad}(\omega) B_0^d e^{i\omega t} + \mathrm{c.c.} \right] \nn  \\
&=& \sum_n f_n {\cal R}_{n\bm{k};L}^{B;a}(t)~.
\eea
In the final line of Eq.~(\ref{Berry_L_mid2}), we expressed the density matrix trace of the unperturbed Berry connection as a single band sum of a band geometric quantity ${\mathcal R}_{n\bm{k};L}^{B;a}(t)$ weighted by the equilibrium Fermi function. This quantity can be interpreted as the orbital part of the magnetic field induced Berry connection correction, 
\bea  \label{Berry_L_final}
{\cal R}_{n\bm{k};L}^{B;a}(t) &=& \tilde{\cal{G}}_{n\bm k;L}^{B;ad}(\omega) B_0^d e^{i\omega t} + \mathrm{c.c.}~,
\eea
where, we defined a band-geometric quantity \emph{orbital}-Magnetic Berry connection polarizability (\emph{orbital}-MBCP) $\tilde{\cal{G}}_{n\bm k;L}^{B;ad}(\omega)$. Its expression is given by, 
\bea  \label{orbital_MBCP}
\tilde{\cal{G}}_{n\bm k;L}^{B;ad}(\omega) &=& {\cal{G}}_{n\bm k;L}^{B;ad}(\omega) + [{\cal{G}}_{n\bm k;L}^{B;ad}(-\omega)]^*~.
\eea
%

%%%%%%%%%%%%%%%%%%%%%%%%%%%%%%%%%%%%%%

\subsubsection{Spin part of the Berry connection correction}

The spin part of the magnetic field driven Berry connection correction can be obtained by evaluating the effective single band expression of the term, 
\bea  \label{Berry_S_int}
\sum_{n,m\neq n} {\rho}_{nm;S}^B(t){\mathcal R}_{mn}^a &=& \sum_{n,m\neq n}{\rho}_{nm;S}^B(\omega){\mathcal R}_{mn}^a ~ e^{i\omega t} + (\omega \rightarrow - \omega)~.
\eea
We substitute the density matrix expression from Eq.~(\ref{rho_B_S}) to obtain
\bea
\sum_{n,m\neq n} \rho_{nm;S}^B (\omega) {\mathcal R}_{mn}^a &=&  \frac{i\mathsf{g}_s \mu_B}{4\hbar} \sum_{n,m\neq n} g_{nm}^{\omega}e^{i \omega t} [\bm{B}_0 \cdot \bm{\sigma}_{nm}]{\mathcal R}_{mn}^a f_{nm} \nn \\
&=& \frac{i\mathsf{g}_s \mu_B}{4\hbar} \sum_{n, m \neq n} e^{i \omega t}\bm{B}_0 \cdot \big[ f_n g_{nm}^{\omega} \bm{\sigma}_{nm} {\mathcal R}_{mn}^a -  f_m g_{nm}^{\omega} \bm{\sigma}_{nm} {\mathcal R}_{mn}^a \big].
\eea
Interchanging the dummy indices $(m \leftrightarrow n)$ for the second term results in, 
\bea   \label{Berry_S_mid}
\sum_{n,m\neq n} \rho_{nm;S}^B (\omega) {\mathcal R}_{mn}^a 
&=& \frac{i\mathsf{g}_s \mu_B}{4\hbar} \sum_{n, m \neq n} e^{i \omega t}\bm{B}_0 \cdot \big[ f_n ~ g_{nm}^{\omega} \bm{\sigma}_{nm} {\mathcal R}_{mn}^a -  f_n ~ g_{mn}^{\omega} \bm{\sigma}_{mn} {\mathcal R}_{nm}^a \big] \nn \\
&=& \frac{i\mathsf{g}_s \mu_B}{4\hbar} \sum_{n, m \neq n} e^{i \omega t}\bm{B}_0 \cdot \big[ f_n ~ g_{nm}^{\omega} \bm{\sigma}_{nm} {\mathcal R}_{mn}^a -  f_n(g_{nm}^{-\omega})^* \bm{\sigma}_{mn} {\mathcal R}_{nm}^a \big]~.
\eea
Substitution of the result from Eq.~(\ref{Berry_S_mid}) into Eq.~(\ref{Berry_S_int}) results in
\bea 
\sum_{n,m\neq n} {\rho}_{nm;S}^B(t){\mathcal R}_{mn}^a &=& \sum_{n,m\neq n}{\rho}_{nm;S}^B(\omega){\mathcal R}_{mn}^a ~ e^{i\omega t} + (\omega \rightarrow - \omega)\nn \\
&=& \frac{i\mathsf{g}_s \mu_B}{4\hbar}    \sum_{n, m \neq n} e^{i \omega t}\bm{B}_0 \cdot \big[ f_n ~ g_{nm}^{\omega} \bm{\sigma}_{nm} {\mathcal R}_{mn}^a -  f_n(g_{nm}^{-\omega})^* \bm{\sigma}_{mn} {\mathcal R}_{nm}^a \big]  + (\omega \rightarrow - \omega)\nn \\
&=& \frac{i}{2\hbar} \sum_{n} f_n \sum_{m \neq n} \cdot \left[ g_{mn}^{\omega} {\cal M}_{mn}^{S;d} {\mathcal R}_{nm}^a - (g_{mn}^{-\omega})^* {\cal M}_{nm}^{S;d} {\mathcal R}_{mn}^a \right] B_0^de^{i \omega t} + \mathrm{c.c.} \nn \\
&=& \sum_n f_n \left[ {\cal{G}}_{n\bm k;S}^{B;ad}(\omega) + [{\cal{G}}_{n\bm k;S}^{B;ad}(-\omega)]^* \right] B_0^de^{i \omega t} + \mathrm{c.c.} \nn \\
&=& \sum_n f_n \left[ \tilde{\cal{G}}_{n\bm k;S}^{B;ad}(\omega) B_0^d e^{i\omega t} + \mathrm{c.c.} \right] \nn  \\
&=& \sum_n f_n {\cal R}_{n\bm{k};S}^{B;a}(t)~.
\eea
To obtain the expression in the third line of the above equation, we have used the relation $(g_{nm}^{-\omega})^* = g_{mn}^{\omega}$. Here, 
\bea  \label{Berry_S_final}
{\cal R}_{n\bm{k};S}^{B;a}(t) &=& \tilde{\cal{G}}_{n\bm k;S}^{B;ad}(\omega) B_0^d e^{i\omega t} + \mathrm{c.c.}
\eea
corresponds to the spin part of the magnetic field induced Berry connection correction. It is related to the \emph{spin}-MBCP tensor
\bea  \label{spin_MBCP}
\tilde{\cal{G}}_{n\bm k;S}^{B;ad} &=&  {\cal{G}}_{n\bm k;S}^{B;ad}(\omega) + [{\cal{G}}_{n\bm k;S}^{B;ad}(-\omega)]^*~,
\eea
where 
\bea
{\cal{G}}_{n\bm k;S}^{B;ad}(\omega) &=& \frac{i}{2\hbar} \sum_{m \neq n} g_{mn}^{\omega} {\cal M}_{mn}^{S;d} {\mathcal R}_{nm}^a~.
\eea
The matrix element of the spin magnetic moment ${\cal M}_{mn}^{S;d} = -(\mathsf{g}_s \mu_B/2)\sigma_{mn}^d$.

Now, we combine the results of orbital and spin contributions to write the total magnetic field induced Berry connection correction, 
\bea  \label{Berry_correction_total}
{\cal R}_{n\bm{k}}^{B;a}(t) &=& {\cal R}_{n\bm{k};L}^{B;a}(t) + {\cal R}_{n\bm{k};S}^{B;a}(t) = \tilde{\cal{G}}_{n\bm{k}}^{B;ad}(\omega)B_0^d e^{i\omega t} + \text{c.c.}
\eea
Here, 
\bea  \label{tot_MBCP}
\tilde{\cal{G}}_{n\bm{k}}^{B;ad}(\omega) = \tilde{\cal{G}}_{n\bm{k};L}^{B;ad}(\omega) + \tilde{\cal{G}}_{n\bm{k};S}^{B;ad}(\omega)~,
\eea
and expressions of $\tilde{\cal{G}}_{n\bm{k};L}^{B;ad}(\omega)$ and $\tilde{\cal{G}}_{n\bm{k};S}^{B;ad}(\omega)$ are given in Eqs. (\ref{orbital_MBCP}) and (\ref{spin_MBCP}), respectively.

%%%%%%%%%%%%%%%%%%%%%%%%%%%%%%%%%%%%%%%%%%%%%%

\subsubsection{Zero frequency result of the Berry connection correction}

In the transport regime $(\omega \tau \ll 1)$ for a dc magnetic field ($\omega = 0$), the Berry connection correction can be split into an intrinsic and extrinsic part. This can be done by approximating the scattering time-dependent $g_{nm}^{\omega}$ term as 
\bea \label{g_nm_approximation}
g_{nm}(\omega = 0) = \frac{1}{\frac{1}{\tau}+i\omega_{nm}} \approx -\frac{i}{\omega_{nm}} + \frac{1}{\omega_{nm}^2 \tau}~.
\eea
For typical topological materials, the scattering time is $\tau \sim 1~\mathrm{ps}$~\cite{Justin_23_prl,ZZDu_19_NatComm}, corresponding to an energy scale $\hbar/\tau \sim 0.7~\mathrm{meV}$. By contrast, in gapped systems, and in gapless systems away from band-touching points, the relevant interband energy scale is typically $\hbar\omega_{nm} \sim 100~\mathrm{meV}$~\cite{XCXie_23_NatRev, Smejkal_17_prl,Atasi_22_NatPhys,Kamal_22_prl}. Consequently, $\omega_{nm}\tau \gg 1$, and the subleading term in the above expansion can be neglected. Keeping only the leading contribution gives $g_{nm}(\omega=0) \simeq -i/\omega_{nm}$, and substituting this approximation into Eq.~(\ref{Berry_correction_total}) yields the magnetic field induced intrinsic Berry-connection correction 
\bea \label{B_induced_BC}
\tilde{{\mathcal R}}_{nn}^{B;a} = -2 \Re\sum_{m\neq n} \frac{\mathcal{R}_{nm}^a [\bm{B} \cdot (\bm{\cal M}_{mn}^L - \bm{\cal M}_{mn}^S]}{\varepsilon_m - \varepsilon_n} - \frac{e}{\hbar} B^d\epsilon_{dbc}(\partial_b \mathrm{g}_n^{ca}) - 2\Re \left[ \sum_{m\neq p} \sum_{p \neq n} \frac{{\mathcal R}_{pm}^a(\bm{B} \cdot \bm{\cal{M}}_{mnp}^L) }{\varepsilon_p - \varepsilon_m} \right]~,\nn \\
\eea
where $\mathrm{g}_n^{ca} = \sum_{m \neq n}\Re(\cal{R}_{nm}^c \cal{R}_{mn}^a)$ is the quantum metric tensor. The last term represents a novel multiband correction to the existing semiclassical expression.

%%%%%%%%%%%%%%%%%%%%%%%%%%%%%%%%%%%%%%%%%%%%%%%%%%%%%%%%%%%%%%%%%%%%%

\section{Equations of motion} \label{Appendix: EOM}

In this section, we aim to derive the equations of motion of Bloch electrons. The dynamics of the charge carriers are governed by the Euler-Lagrange equations 
\bea \label{k_dot}
\quad \frac{d}{dt} \left( \frac{\partial \mathcal{L}_{n \bm k}}{\partial \dot{\bm{\mathfrak{r}}}_{n}} \right) &=& \frac{\partial \mathcal{L}_{n \bm k}}{\partial \bm{\mathfrak{r}}_{n}}~, \\
\frac{d}{dt} \left(\frac{\partial \mathcal{L}_{n \bm k}}{\partial \dot{\bm{k}}} \right) &=& \frac{\partial \mathcal{L}_{n \bm k}}{\partial \bm{k}}~.
\eea
For convenience, we write the Lagrangian expression derived in Eq.~(\ref{Lag_B_t})
\bea 
\mathcal{L}_{n\bm k} &=& - \hbar\dot{\bm{k}} \cdot \left[\bm{\mathfrak{r}}_n - \left( \bm{\mathcal R}_{n\bm{k}} + \bm{\mathcal{R}}_{n\bm{k}}^{B}(t)\right)\right] + \frac{e}{2}[\bm{B}(t)\times \dot{\bm{\mathfrak r}}_n] \cdot \bm{\mathfrak{r}}_n - \tilde{\varepsilon}_{n\bm k}~.
\eea
The first Euler-Lagrange equation yields the force equation, 
\bea
&& \qquad \frac{e}{2}[\dot{\bm{\mathfrak{r}}}_n \times \bm{B}(t)] = - \hbar \dot{\bm{k}} + \frac{e}{2}[\bm{B}(t) \times \dot{\bm{\mathfrak{r}}}_n] \nn \\
&& \Rightarrow \hbar \dot{\bm{k}} = -e[\dot{\bm{\mathfrak{r}}}_n \times \bm{B}(t)]~,\label{force_equation}
\eea
The second Euler-Lagrange equation produces the velocity equation of the charge particle,
\bea
\frac{d\mathfrak{r}_n^a}{dt} &=& \frac{1}{\hbar}\frac{\partial \tilde{\varepsilon}_{n\bm{k}}}{\partial k_a} + \left( \frac{d \cal{R}_{n\bm{k}}^{a}}{dt} + \frac{d \cal{R}_{n\bm{k}}^{B;a}}{dt} \right) - \frac{\partial}{\partial k_a}\left[ \dot{\bm{k}} \cdot \left( \bm{\mathcal R}_{n\bm{k}} + \bm{\mathcal{R}}_{n\bm{k}}^{B}(t)\right) \right] \nn \\
\Rightarrow \dot{\mathfrak{r}}_{n}^a &=& \tilde{v}_{n\bm{k}}^a + \left( \dot{\bm{k}} \cdot \frac{\partial {\mathcal R}_{nn}^a}{\partial \bm{k} } -  \dot{\bm{k}} \cdot \frac{\partial \bm{\mathcal R}_{n\bm{k}}}{\partial k_a} \right) + \left(   \dot{\bm{k}} \cdot \frac{\partial {\mathcal R}_{n\bm{k}}^{B;a}}{\partial \bm{k}} -  \dot{\bm{k}} \cdot \frac{\partial \bm{\mathcal R}_{n\bm{k}}^{B;a}}{\partial k_a} \right) + \frac{\partial \bm{\mathcal R}_{n\bm{k}}^{B;a}}{\partial t} \nn \\
&=& \tilde{v}_{n\bm{k}}^a - [\dot{\bm{k}} \times ( \bm{\Omega}_{n\bm{k}} + \bm{\Omega}_{n\bm{k}}^B )]^a + \frac{\partial \bm{\mathcal R}_{n\bm{k}}^{B;a}}{\partial t}~.\label{velocity_equation}
\eea
Here, $\tilde{v}_{n\bm k}^a$ is the magnetic field corrected band velocity and $ \bm{\Omega}_{n\bm k} ( \bm{\Omega}_{n\bm k}^B) = \partial_{\bm {k}}\times \bm{\cal R}_{n\bm k} (\bm{\cal R}_{n \bm k}^B)$ is the intrinsic (magnetic field induced) Berry curvature of the system. The second term produces an anomalous velocity of the charge carrier. The third term is a consequence of an oscillating magnetic field and would vanish in the dc limit. The gyrotropic current due to the spin part of the oscillating Berry connection has been demonstrated in a recent article by Jian Wang \emph{et al.}~\cite{Wang_2025_prl}. Recently, a similar velocity component has been discovered in case of an oscillating electric field, yielding the electric displacement current~\cite{Jian_Wang_2024_PRB}, and quantum capacitance in insulators~\cite{Tobias_2024_NatComm}. Eqs. (\ref{force_equation}) and (\ref{velocity_equation}) can be decoupled and written in the following form
\bea
\dot{\bm{\mathfrak{r}}}_{n\bm{k}} &=& D_{n \bm{k}}\left[ \tilde{\bm v}_{n\bm{k}} + \frac{e}{\hbar}(\tilde{\bm v}_{n\bm{k}} \cdot \tilde{\bm{\Omega}}_{n\bm{k}})\bm{B} + \frac{\partial \bm{\cal R}_{n\bm{k}}^{B}}{\partial t} + \frac{e}{\hbar}\left( \frac{\partial \bm{\cal R}_{n\bm{k}}^{B}}{\partial t} \cdot \tilde{\bm{\Omega}}_{n\bm{k}}\right) \bm{B} \right]~\label{EOM_r_total},\\
\hbar \dot{\bm{k}} &=& D_{n\bm{k}}\left[ -e(\tilde{\bm v}_{n\bm{k}} \times \bm{B}) - e\left( \frac{\partial \bm{\cal R}_{n\bm{k}}^{B}}{\partial t} \times \bm{B} \right)\right] ~, \label{EOM_k_app}
\eea
where $\tilde{\bm{\Omega}}_{n\bm{k}} = ( \bm{\Omega}_{n\bm{k}} + \bm{\Omega}_{n\bm{k}}^B )$ is total Berry curvature.

%%%%%%%%%%%%%%%%%%%%%%%%%%%%%%%%%%%%%%%%%%%%%%%%%%%%%%%%%%%%%%%%%%%%%%%

\section{Intrinsic gyrotropic magnetic current} \label{Appendix: IGMC} 

In the last section, we derived the particle velocity in the presence of an oscillating magnetic field. This velocity weighted by the momentum space equilibrium Fermi function yields the intrinsic charge current, 
\be
    \bm{j} = -e\int_{n{\bm k}} D_{n{\bm k}}^{-1}\dot{\bm{\mathfrak{r}}}_{n\bm {k}} f_{n\bm {k}}~,
\ee
where, we have defined $\int_{n\bm k}\equiv \sum_n d^d\bm{k}/ (2 \pi)^d$, $d$ is the spatial dimension. Substitution of the result from Eq.~(\ref{EOM_r_total}) into the above equation and keeping terms only up to first order in the magnetic field, we obtain the linear charge current to be 
\bea
 \bm{j} = -e\int_{n\bm {k}} \left[ {\bm v}_{n\bm{k}}^B + \frac{e}{\hbar}({\bm v}_{n\bm{k}} \cdot {\bm{\Omega}}_{n\bm{k}})\bm{B} + \frac{\partial \bm{\cal R}_{n\bm{k}}^{B}}{\partial t} \right]f_{n\bm k}~.
\eea
%.
Now, as shown in the main text, the oscillating magnetic field induced linear charge current can be written as
\begin{equation}
    j^{\text{GMC}}_a (t) = \chi^{\text{GMC}}_{a;d} (\omega) B^d_0 e^{i \omega t} + \mathrm{c.c.}
\end{equation}
Here, $\chi^{\text{GMC}}_{a;d} (\omega) $ is the gyrotropic magnetic conductivity for the current flowing along $a-$axis with the field along $d-$axis. Depending on physical origins, the response tensor $ \chi^{\text{GMC}}_{a;d} (\omega) $ can be decomposed into three different parts, which are Fermi surface oscillation (FO), chiral magnetic velocity (CMV), and dynamic charge polarization, denoted as displacement current (Disp), i.e.,
\begin{equation}
    \chi^{\text{GMC}}_{a;d} (\omega) = \chi_{a;d}^{\mathrm{G,FO}} + \chi^{\text{G,Disp}}_{a;d} +  \chi^{\text{G,CMV}}_{a;d}~.
\end{equation} 
The explicit expressions of $ \chi^{\text{G, FO}}_{a;d} (\omega),~\chi^{\text{G, CMV}}_{a;d} (\omega),~\text{and}~ \chi^{\text{G, Disp}}_{a;d} (\omega)$ are
\bea
\chi_{a;d}^{\mathrm{G,FO}}(\omega) &=&  -ie\omega g^{\omega}_0 \int_{n\bm k} v_{n\bm{k}}^a {\cal M}_{n\bm{k}}^d \frac{\partial f_{n\bm{k}}}{\partial \varepsilon_{n\bm{k}}}~,\label{Fermi_oscillation_app} \\
\chi_{a;d}^{\mathrm{G,Disp}}(\omega) &=& -ie\omega \int_{n\bm k}
\tilde{\mathcal G}_{n\bm k}^{B;ad}(\omega) f_{n\bm k}~,\label{Displacement_contribution_app} \\
\chi_{a;d}^{\mathrm{G,CMV}}(\omega) &=& -\frac{e^2}{\hbar}\delta_{ad}\int_{n\bm k}({\bm v}_{n\bm{k}} \cdot {\bm{\Omega}}_{n\bm{k}})f_{n\bm k}~.\label{chiral_magnetic_velocity_app}
\eea
From Eq.~\eqref{chiral_magnetic_velocity_app}, it is clear that the chiral magnetic velocity contribution is purely intrinsic and driven by the Berry curvature of the system. But, the other two contributions of GMC, namely the conventional Fermi surface term and the displacement currents, depend on the scattering time $\tau$. Their $\tau$ dependence originate from the factors $g_0^{\omega} = [1/\tau + i\omega]^{-1} $ and $ g^{\omega}_{nm} = [1/\tau + i (\omega_{nm} + \omega) ]^{-1}$. These factors can be expanded binomially, and based on the relative magnitudes of the scattering time, driving frequency and characteristic interband transition frequency, they can be segregated into an intrinsic and extrinsic part. Here, we work in the transport regime, where $\omega \tau \ll 1$. In this limit, these factors approximate as $g_0^{\omega} \approx \tau$ and $ g^{\omega}_{nm} \approx -i/\omega_{nm}$ [See Eq. (\ref{g_nm_approximation})]. Substituting these results into Eqs. (\ref{Fermi_oscillation_app}) and (\ref{Displacement_contribution_app}) yields the conventional GMC to be purely extrinsic and the displacement contribution a purely intrinsic response. Thereby, the intrinsic gyrotropic magnetic current (IGMC) is given by
\bea  \label{chi_IGMC_app}
\chi_{a;d}^{\mathrm{IGMC}}(\omega) = \chi_{a;d}^{\mathrm{IG,Disp}} + \chi_{a;d}^{\mathrm{IG,CMV}}~.
\eea
Here, the displacement contribution of IGMC depends on the intrinsic part of the MBCP tensor in the transport regime, given by
\bea
\tilde{G}_{n \bm k; L/S}^{B;ad} &=& \lim_{\omega \ll 1/\tau} \tilde{\mathcal{G}}_{n \bm k; L/S}^{B;ad} (\omega)  = {G}^{B;ad}_{n \bm k; L/S} + [{G}^{B;ad}_{n \bm k; L/S} ]^*~,
\eea
where
\bea
{{G}}_{n\bm k;L}^{B;ad} &=& -\frac{e}{2\hbar}\sum_{m\neq n} \epsilon_{dbc} (\partial_b G_n^{ca}) -\Re\sum_{m\neq n} \frac{\cal{M}_{mn}^{L;d}{\mathcal R}_{nm}^a}{\varepsilon_m - \varepsilon_n} + \Re \sum_{m\neq p} \sum_{p \neq n} \frac{\cal{M}_{mnp}^{L;d}{\mathcal R}_{pm}^a}{\varepsilon_m - \varepsilon_p}~, \\
\tilde{{G}}_{n \bm k;S}^{B;ad}  &=& \Re \sum_{m \neq n}\frac{{\cal M}_{mn}^{S;d} {\mathcal R}_{nm}^a}{\varepsilon_m - \varepsilon_n}~.
\eea
Importantly, in the low frequency regime, the driving frequency being much lower compared to the system's interband frequency, the band geometric MBCP tensors become independent of the driving frequency and coincide with the zero frequency result [see Eq. (\ref{B_induced_BC})].

%%%%%%%%%%%%%%%%%%%%%%%%%%%%%%%%%%%%%%%%%%%%%%%%%%%%%%%%%%%%%%%%%%%%%%%

\section{Magnetic point group symmetry of the IGMC tensors} \label{Sec: mag_point_group}

The IGMC response tensor can be segregated into $\mathcal T$-even and $\mathcal T$-odd parts based on the time-reversal parity of the corresponding conductivity expressions. A microscopic symmetry analysis of the two contributing channels: displacement current and chiral magnetic velocity, shows that $\chi_{a;d}^{\mathrm{IG,CMV}}$ is $\mathcal T$-even and can therefore be finite in both magnetic and nonmagnetic systems. In contrast, the displacement contribution $\chi_{a;d}^{\mathrm{IG,Disp}}$ is $\mathcal T$-odd and vanishes in nonmagnetic materials. To assess the feasibility of these contributions for nonmagnetic systems, we presented the crystallographic point group analysis in Table~\ref{table_nonmag_point_group} of Sec .~\ref {Sec: Symmetry}. We now complete the symmetry classification by examining the symmetry for magnetic point groups. Notably, the symmetry tables reveal that both these contributions vanish identically in systems possessing $\mathcal C_{3,4,6}^{z}\mathcal T$ symmetry, or equivalently pure rotational symmetries $\mathcal C_{2,3,4,6}^{z}$.

%%%%%%%%%%%%%%%%%%%%%%%%%%%%%%%%%%%%%%%%%%%%%%

\begingroup
\setlength{\tabcolsep}{3 pt}
\begin{table*}[b!]
\caption{ The symmetry restrictions of the magnetic gyrotropic response tensors. The cross (\xmark) and the tick (\cmark) mark signify that the corresponding response tensor is symmetry forbidden and allowed, respectively. Here, ${\cal M}_{a}$, ${\cal C}_{n}^a$ and ${\cal S}_{n}^a$ represent  mirror, $n$-fold rotation, and $n$-fold roto-reflection symmetry operation along the $a$-direction for $a=\{x,y,z\}$, respectively. }
\begin{tabular}{c c c c c c c c c c c c c c c c c c}
\hline \hline 
\noalign{\vskip 2pt}
IGMC & ${\cal M}_x\cal{T}$ & ${\cal M}_y\cal{T}$ & ${\cal M}_z\cal{T}$ & ${\cal C}_{2}^x\cal{T}$ & ${\cal C}_{2}^y\cal{T}$ & ${\cal C}_{2}^z\cal{T}$ & ${\cal C}_{4}^x\cal{T}$ & ${\cal C}_{4}^y\cal{T}$ & ${\cal C}_{3,6}^x\cal{T}$ & ${\cal C}_{3,6}^y\cal{T}$ & ${\cal C}_{3,4,6}^z\cal{T}$ & ${\cal S}_{4}^x\cal{T}$  & ${\cal S}_{6}^x\cal{T}$ & ${\cal S}_{4}^y\cal{T}$  & ${\cal S}_{6}^y\cal{T}$ & ${\cal S}_{4,6}^z\cal{T}$  \\
\noalign{\vskip 6pt}
\hline \hline 

\noalign{\vskip 6pt}

$\chi_{x;z}^{\mathrm{IG,Disp}}$ & \xmark  & \cmark & \xmark & \cmark & \xmark & \cmark & \xmark & \cmark & \xmark & \xmark & \xmark & \xmark & \xmark & \cmark & \cmark & \xmark  \\

\noalign{\vskip 6pt}

$\chi_{y;z}^{\mathrm{IG,Disp}}$  & \cmark & \xmark & \xmark & \xmark & \cmark & \cmark & \cmark & \xmark & \xmark & \xmark & \xmark & \cmark & \cmark & \xmark & \xmark & \xmark  \\

\noalign{\vskip 6pt} 

\hline

\noalign{\vskip 6pt} 

$\chi_{x;z}^{\mathrm{IG,CMV}}$ & \cmark & \xmark & \cmark & \xmark & \cmark & \xmark & \xmark & \cmark  & \xmark & \cmark & \xmark & \xmark & \xmark & \cmark & \xmark & \xmark  \\

\noalign{\vskip 6pt}

$\chi_{y;z}^{\mathrm{IG,CMV}}$ & \xmark & \cmark & \cmark & \cmark & \xmark & \xmark & \cmark & \xmark & \cmark & \xmark & \xmark & \cmark & \xmark & \xmark & \xmark & \xmark  \\

\noalign{\vskip 6pt}
\hline \hline
\end{tabular}
\label{table_mag_point_group}
\end{table*}
\endgroup

%%%%%%%%%%%%%%%%%%%%%%%%%%%%%%%%%%%%%%%%%%%%%%%%%%%%%%%%%%%%%%%%%%%%%%%%%%%%%

\twocolumngrid
\bibliography{Refs}
\end{document}